\documentclass[preprint,12pt]{elsarticle}

\usepackage[hang]{footmisc}
\usepackage{amssymb}
\usepackage{bm}
\usepackage{psfrag}
\usepackage{amsthm,color}

\usepackage[lined,boxed,commentsnumbered]{algorithm2e}

\usepackage{epsf,amsfonts,amsmath}

\newtheorem{prop}{Proposition}
\newtheorem{rem}{Remark}

\journal{Elsevier}

\begin{document}

\begin{frontmatter}

\title{Non-intrusive Low-Rank Separated Approximation of High-Dimensional Stochastic Models}

\author[UCB1]{Alireza Doostan\corref{cor1}}
\ead{doostan@colorado.edu}

\author[UCB2]{AbdoulAhad Validi}
\ead{AbdoulAhad.Validi@colorado.edu}

\author[Stan]{Gianluca Iaccarino}
\ead{jops@stanford.edu}

\cortext[cor1]{Corresponding Author: Alireza Doostan}

\address[UCB1]{Aerospace Engineering Science Department,University of Colorado, Boulder, CO 80309, USA}
\address[UCB2]{Mechanical Engineering Department, University of Colorado, Boulder, CO 80309,  USA}
\address[Stan]{Department of Mechanical Engineering, Stanford University, Stanford, CA 94305,  USA}

\begin{abstract}
\label{Abstract}

This work proposes a sampling-based (non-intrusive) approach within the context of {\it low-rank separated representations} to tackle the issue of curse-of-dimensionality associated with the solution of models, e.g., PDEs/ODEs, with high-dimensional random inputs. Under some conditions discussed in details, the number of random realizations of the solution, required for a successful approximation, grows linearly with respect to the number of random inputs. The construction of the separated representation is achieved via a regularized alternating least-squares regression, together with an error indicator to estimate model parameters. The computational complexity of such a construction is quadratic in the number of random inputs. The performance of the method is investigated through its application to three numerical examples including two ODE problems with high-dimensional random inputs.

\end{abstract}


\end{frontmatter}


%
\section{Introduction}
\label{introduction}

In most practical situations, the behavior and evolution of complex systems are known only with limited certainty. This is mainly due to the lack of knowledge about the governing physical laws or limited information regarding their operating conditions and input parameters, e.g., fuel or material properties. A reliable computer simulation of such systems, therefore, requires a systematic representation of uncertainties and quantification of their impact on quantities of interest. 

In the probabilistic framework, uncertainties are represented using random variables and processes characterized by, for instance, the available experimental data or expert opinion. A major task is then to quantify the dependence of quantities of interest -- also random variables or processes -- on these random inputs. Development of efficient numerical tools for the computation of such mappings has been a research subject over the past few decades and has become an emerging field of study more recently. 

Statistical tools such as Monte Carlo sampling and its variations have been widely used for this purpose. To address the low convergence rate of these methods, stochastic basis expansions, for instance, in chaos polynomials \cite{Ghanem03,Ghanem99b,Xiu02} and multivariate numerical integration/interpolation, such as those based on the sparse grid collocation \cite{Tatang95,Mathelin03,Xiu05a}, have been recently proposed. While proven efficient both numerically and analytically on numerous problems in engineering and science, these techniques may face difficulties when input uncertainties are characterized by a large number of independent random variables (i.e., high-dimensional random inputs) \cite{Xiu05a,Xiu10a}. Specifically, the computational complexity of these methods, in their original forms, grows exponentially fast with respect to the number of input random variables: an issue known as the {\it curse-of-dimensionality.} The reason for such a fast growth is the tensor product construction of multi-dimensional bases (in polynomial chaos methods) or quadrature rules (in sparse grid collocation approaches) from one-dimensional bases or quadrature rules, respectively. Such tensorizations, therefore, impose an explicit dependence on the input dimensionality. Instead, a number of recent techniques have been proposed that exploit, for instance,  {\it low-rank} or {\it sparsity}  structures of quantities of interest in order to reduce this exponential complexity growth, see, e.g., \cite{Doostan07,Todor07a,Nobile08b,Bieri09a,Ma09a,Bieri09b,Blatman10,Doostan11a,Nouy07,Doostan09}. However, effective treatment of the curse-of-dimensionality still remains an open problem.

 
Alternatively, in the present study, we propose an approach that constructs a {\it separated representation} of the solution of interest. Specifically, let $u(\bm y)$ with $\bm{y}(\omega)=\left(y_1(\omega),\ldots, y_d(\omega)\right):\Omega\rightarrow \mathbb{R}^{d}$, $\omega\in\Omega$ and $d\in\mathbb{N}$, be a stochastic function defined on a suitable probability space $(\Omega,\mathcal{F},\mathcal{P})$. Then $u(\bm y)$ admits a separated representation if, for some (small) $r\in\mathbb{N}$ and accuracy $\epsilon\ge 0$,
\begin{equation}
\label{eq:sep_rep_intro}
u(\bm y)=\sum_{l=1}^{r}s_{l}u_{1}^{l}(y_1)\cdots u_d^{l}(y_d)+\mathcal{O}(\epsilon),
\end{equation}
where the univariate functions $\{u_k^l(y_k)\}_{k=1}^d$, $l=1,\dots,r$, are unknown and $\{s_l\}$, $l=1,\dots,r$, are some normalization constants. When the {\it separation rank} $r$ is independent of $d$, the approximation (\ref{eq:sep_rep_intro}) may be obtained with a computational complexity that is linear in $d$, \cite{Beylkin02,Beylkin05,Beylkin08,Doostan09}, hence drastically reducing the curse-of-dimensionality. When the approximation (\ref{eq:sep_rep_intro}) is achieved by a small $r$, then $u(\bm y)$ is said to admit a {\it low-rank separated representation}. 

Separated representations, also known as canonical decompositions (CANDECOMP) or parallel factor analysis (PARAFAC), have been first introduced by Hitchcock \cite{Hitchcock} to represent a multi-way tensor as a finite sum of rank-one tensors. Consequently, they have been extensively applied to several areas including image compression and classification \cite{Shashua01,Furukawa02}, telecommunication \cite{Sidiropoulos00,Sidiropoulos02,Lathauwer07}, neuroscience \cite{Mocks1988,Field1991,Andersen04}, chemometrics \cite{Bro97a}, and data mining \cite{Beylkin02,Beylkin05,Kroonenberg80,Kolda01,Tyrtyshnikov04,Hackbusch04,Acar05,Beylkin08,Kolda09a,Kolda09b}. 

Separated representations have been recently used for the reduced order solution of deterministic PDEs in solid and fluid mechanics, where they are often called proper generalized decompositions (PGDs), see, e.g., \cite{Ladeveze89,Ladeveze99,Ammar06,Ammar07,Chinesta10a,Neron10,Chinesta10b,Ladeveze11a,Chinesta11a,Chinesta11b} and the references therein. 

In the context of stochastic problems, an expansion based on the separation of spatial and random variables has been proposed in \cite{Nouy07,Nouy08}. While this approach significantly reduces the computational complexity of full-order polynomial chaos (PC) approximations, it suffers from the curse-of-dimensionality for high-dimensional random inputs. A separated representation of the form (\ref{eq:sep_rep_intro}) for the solution of PDEs and ODEs with high-dimensional random inputs has been first studied in \cite{Doostan07b,Doostan08b,Doostan09}. Similar representations, although with different solution strategies, have been later investigated in \cite{Nouy10,Khoromskij10,Falco11}. 

The current separated approximation of models with high-dimensional random inputs are {\it intrusive}, that is, one has to rewrite deterministic solvers in order to propagate the uncertainty. However, for many large-scale complex systems, it is desirable to develop {\it non-intrusive} solvers, where deterministic codes are treated as black box. Such a construction is the purpose of the present study. Specifically, we here extend the alternating least-squares regression approach of \cite{Beylkin08} to problems with high-dimensional random inputs. To enhance the stability and accuracy of this technique, we propose a Tikhonov regularization of the regression problem along with an error indicator for the selection of two main parameters of the separated representation. Under some conditions to be discussed in Section \ref{sec:cost}, the number of random realizations of the solution, required for a successful approximation, grows linearly with respect to $d$. Furthermore, the computational complexity of the alternating least-squares regression approach is quadratic in $d$.

The rest of this paper is structured as follows. In Section \ref{problem_setup}, we introduce the problem of interest in an abstract form. Following that, in Section \ref{sec:Separated Representation}, we discuss the separated representations in more details and describe their non-intrusive construction using an alternating least-squares regression. Additionally, we discuss the computational complexity of this non-intrusive method. In Section \ref{sec:overfitting}, we propose a regularization approach together with an error indicator to derive a stopping criterion and, hence, avoid over-fitting the solution realizations. Finally, in Section \ref{sec:num_exp}, we demonstrate the performance and efficiency of the proposed approach on a manufactured function as well as two ODE problems with high-dimensional random inputs. The first ODE problem is a linear elliptic equation with random diffusion coefficient. The second one is concerned with a hydrogen oxidation problem where reaction rate constants and species thermodynamics are uncertain.

\section{Problem setup}
\label{problem_setup}

Let $\left( \Omega,\mathcal{F},\mathcal{P} \right)$ be a complete probability space on which the random inputs are defined. Here $\Omega$ is the set of elementary events, $\mathcal{F}\subset 2^{\Omega}$ is the $\sigma$-algebra of events, and $\mathcal{P}:\mathcal{F}\rightarrow [0,1]$ is a probability measure function defined on $\mathcal{F}$. For the sake of demonstration, we consider a generic stochastic ordinary differential equation (ODE)
\begin{equation} \label{eq:system of ODE}
\mathcal{A} \left(t,\bm{y}(\omega)  ;u \right)=0,\ \ \left(t, \omega\right) \in \left[0,T\right]\times\Omega,
\end{equation}
that holds $\mathcal{P}$-{\it a.s.} for $\omega\in\Omega$. Appropriate initial condition and forcing function are also considered. Here, $\mathcal{A}$ denotes the ordinary differential operator, $t\in[0,T],\; T>0$, is the time variable, and $\bm{y}(\omega)=\left(y_1(\omega),\ldots, y_d(\omega)\right):\Omega\rightarrow \mathbb{R}^{d}$, $d\in\mathbb{N}$, represents the vector of random variables defining the input uncertainty. We further assume that the components of the random vector $\bm{y}(\omega)$ are statistically independent and identically distributed (i.i.d.) according to a probability density function $\rho(y_k):\Gamma\subseteq\mathbb{R} \rightarrow\mathbb{R}_{\ge 0}$, $k=1,\dots,d$.

Our goal is to estimate the possibly non-linear solution
\begin{equation}
\label{eq:solution}
u(t,\bm y):=u(t,\bm y(\omega)) = u(t,y_1(\omega),\dots,y_d(\omega)): \left[0,T\right]\times \Gamma^d\rightarrow \mathbb{R}
\end{equation}
of (\ref{eq:system of ODE}) in a non-intrusive fashion, i.e., with the assumption that the deterministic solver of (\ref{eq:system of ODE}) is a black box. This entails the evaluation of $u(t,\bm y)$ at a set of random or deterministic realizations of $\bm y$ and subsequently the construction of an approximation $\hat{u}(t,\bm y)$ based on an interpolation, a regression, or a discrete projection scheme. 

In the present study, we consider the random sampling of $\bm y$, similar to the standard Monte Carlo simulation. We denote by $D$ a set of $N$ independent realizations $\bm y^{(j)}$, $j=1,\dots,N$, drawn according to the probability density function $\rho(\bm{y})=\prod_{k=1}^{d}\rho(y_k)$, and the corresponding solution realizations $u(t,\bm y^{(j)})$,
\begin{equation}
\label{eq:set of scattered data}
D=\left\lbrace\left( \bm{y}^{(j)}; u(t,\bm{y}^{(j)})  \right)\right\rbrace_{j=1}^N.
\end{equation}

Using the {\it data} set $D$, we then extend the regression approach of \cite{Beylkin08} to stably construct $\hat{u}(t,\bm y)$ in the separated form (\ref{eq:sep_rep_intro}).  

For the interest of description and without loss of generality, we henceforth restrict our analyses to a fixed instance of $t$ and adopt the short notation $u(\bm y)$ for the solution of interest. We next introduce the separated representation (\ref{eq:sep_rep_intro}) in more details and subsequently discuss our non-intrusive approach to construct and regularize such an approximation.


%
\section{Separated representations}
\label{sec:Separated Representation}
%


To motivate the separated representation of $u(\bm y)$, we first briefly describe the standard multivariate approximation of $u(\bm y)$ using polynomial chaos (PC) expansions \cite{Ghanem03,Xiu02}. 

Let $\{\psi_{\alpha_k}(y_k)\}_{k=1}^d$ be the set of univariate spectral polynomials of degree $\alpha_k\in\mathbb{N}_0:=\mathbb{N}\cup\{0\}$ orthonormal with respect to the probability density function $\rho(y_k)$, i.e.,
\begin{equation}
\label{eq:pce_uni} 
\int_{\Gamma} \psi_{\alpha_k}(y_k) \psi_{\beta_k}(y_k) \rho(y_k) d y_k = \delta_{\alpha_k\beta_k}.\nonumber
\end{equation}

A tensorization of the univariate basis $\{\psi_{\alpha_k}(y_k)\}_{k=1}^d$, i.e.,
\begin{equation}
\label{eq:multi_pc_basis} 
\psi_{\bm\alpha}(\bm y) = \psi_{\alpha_1}(y_1)\psi_{\alpha_2}(y_2)\dots\psi_{\alpha_d}(y_d),\quad \bm\alpha\in\mathbb{N}_0^d,
\end{equation}
with $\mathbb{N}_0^d=\{(\alpha_1,\dots,\alpha_d):\;\;\alpha_k\in\mathbb{N}_0\}$, forms an orthonormal basis in $L_2(\Gamma^d,\mathcal{P})$. Then, for $u(\bm y)\in L_2(\Gamma^d,\mathcal{P})$, the infinite series 
\begin{equation}
\label{eq:pce_multi} 
u(\bm y)= \sum_{\bm\alpha\in\mathbb{N}_0^d} s_{\bm \alpha} \psi_{\bm\alpha}(\bm y)
\end{equation}
with coefficients 
\begin{equation}
\label{eq:pce_coef} 
s_{\bm\alpha}= \int_{\Gamma^d} u(\bm y) \psi_{\bm\alpha}(\bm y)\rho(\bm y)d\bm y
\end{equation}
converges in the mean-squares sense. The series (\ref{eq:pce_multi}) is referred to as the polynomial chaos (PC) expansion. 

We highlight a number of properties of the PC representation in (\ref{eq:pce_multi}). Firstly, as observed in (\ref{eq:multi_pc_basis}), the multivariate basis functions $\psi_{\bm\alpha}(\bm y)$ are of separated form with respect to random inputs $y_k$. Secondly, the basis functions $\psi_{\bm\alpha}(\bm y)$ are {\it a priori} selected based on the probability density function $\rho(y_k)$ of $y_k$. Therefore, the PC expansion (\ref{eq:pce_multi}) has a linear structure with respect to the coefficients $s_{\bm\alpha}$. Thirdly, due to the particular tensor product construction (\ref{eq:multi_pc_basis}) of $\psi_{\bm\alpha}(\bm y)$, the cardinality of a finite degree basis $\{\psi_{\bm\alpha}(\bm y)\}$ with, for instance, $\bm\alpha\in\mathbb{N}_0^{d,p}=\{\bm{\alpha}\in\mathbb{N}_0^d:\;\; \Vert\bm\alpha\Vert_1\le p,\ p\in\mathbb{N}_0\}$ grows exponentially with $d$. This fast growth leads to the issue of curse-of-dimensionality as the computational complexity of estimating the coefficients $s_{\bm\alpha}$ grows exponentially fast.

Alternatively, the separated representation of a function $u(\bm y)$ is a decomposition of the form
\begin{equation}
\label{eq:separated_intro}
u(\bm y)=\sum_{l=1}^{r}s_{l}u_{1}^{l}(y_1)\cdots u_d^{l}(y_d)+\varepsilon(\bm y).
\end{equation}

Here, the {\it separation rank} $r$ and univariate {\it factors}/functions $\{u_{k}^{l}(y_k)\}_{k=1}^{d}$, $l=1,\dots,r$, are not fixed {\it a priori} and are outcomes of the approximation such that a prescribed accuracy $\Vert \varepsilon\Vert\le \epsilon$ is achieved. The scalars $\{s_l\}$, $l=1,\dots,r$, may be viewed as some positive normalization constants which we will specify later. We note that, by construction, the separated representation (\ref{eq:separated_intro}) is a non-linear expansion and when $u$ admits a low separation rank $r$, a fast decay of the error with respect to $r$ may be achieved. Due to its separated form with respect to the variables $y_k$, the non-linear approximation (\ref{eq:separated_intro}) may be computed using multi-linear approaches such as the alternating least-squares (ALS) scheme described in Section \ref{sec:als}. As we shall see in Section \ref{sec:cost}, for situations where $r$ is independent of $d$, the separated representation (\ref{eq:separated_intro}) may be obtained with a number of samples $N$ and computational cost that are, respectively, {\it linear} and {\it quadratic} in $d$. In practice, however, if $r$ weakly depends on $d$, e.g., $r\sim \mathcal{O}(\log d)$, the above dependencies will be near-linear and near-quadratic, respectively.

\begin{rem}[Connection with sparse PC expansions] Due to the tensor product construction of PC basis functions $\psi_{\bm\alpha}(\bm y)$ in (\ref{eq:multi_pc_basis}), a {\it sparse} PC expansion is of a low-rank separated form. By sparse PC expansions, we mean representations of the form (\ref{eq:pce_multi}) in which many of the coefficients $s_{\bm\alpha}$, $\bm\alpha\in\mathbb{N}_0^{p,d}$, are negligible. The relevance of the sparse PC approximation of PDEs with random inputs has been discussed in several work including \cite{Bieri09a,Cohen10a,Doostan11a}.
\end{rem}

\subsection{A non-intrusive construction}
\label{sec:non-intrusive}

Given the realizations $\{\bm y^{(j)}\}$ in $D$, we formulate a discrete (pseudo) inner product between $u(\bm y),v(\bm y):\Gamma^d\rightarrow\mathbb{R}$ as 
\begin{equation}
\label{eq:inner product}
\left\langle u,v\right\rangle_D = \left\langle \lbrace \bm{y}^{(j)},u(\bm{y}^{(j)}) \rbrace_{j=1}^N, \lbrace \bm{y}^{(j)},v(\bm{y}^{(j)}) \rbrace_{j=1}^N \right\rangle_D = \frac{1}{N}\sum_{j=1}^N u(\bm{y}^{(j)}) v(\bm{y}^{(j)}),\nonumber
\end{equation}
which induces the (pseudo) norm
\begin{equation}
\label{eq:psuedo_norm}
\Vert u \Vert_D = \left\langle u,u\right\rangle_D^{1/2}.
\end{equation}

Let
\begin{equation}
\label{eq:separated_space}
\mathcal{U}_r= \left\{\sum_{l=1}^{r}s_{l}u_{1}^{l}(y_1)\cdots u_d^{l}(y_d):\; u_k^{l}(y_k):\Gamma\rightarrow\mathbb{R}, s_l\in \mathbb{R}_{>0}, \ \begin{array}{c}k=1,\dots,d \\l=1,\dots,r\end{array}\right\} \nonumber
\end{equation}
be the space of separated real functions in $d$ dimensions and with separation rank $r$. The non-intrusive separated approximation of $u(\bm y)$ may then be cast into a least-squares regression problem of the form
\begin{equation} 
\label{eq:low-rank_regression}
u_r = \arg\min_{\hat{u}_r\in\mathcal{U}_r}\ \Vert u - \hat{u}_r\Vert_D^2.
\end{equation}

Due to the tensor product structure of $\mathcal{U}_r$, the optimization problem (\ref{eq:low-rank_regression}) is non-linear. Furthermore, the separation rank $r$ is not known {\it a priori} and has to be estimated. A discrete formulation of (\ref{eq:low-rank_regression}), as we shall describe in Section \ref{sec:als}, may be solved using non-linear programming techniques such as the damped Gauss-Newton, see, e.g., \cite{Paatero1997}. These non-linear schemes, however, suffer from the curse-of-dimensionality and their current use is thus limited to only low-dimensional problems. Alternatively, we here adopt a multi-linear optimization technique known as the alternating least-squares (ALS) to solve (\ref{eq:low-rank_regression}). 
We next describe the details of such an approach. 

\subsection{Alternating least-squares (ALS) approach}
\label{sec:als}

In the ALS method, we construct a sequence of one-dimensional least-squares regression problems to solve for the univariate functions $\{u_k^{l}(y_k)\}$, $l=1,\dots,r$, along a direction $k$. In doing so, we freeze variables along all other directions at their current approximation. We then alternate over all directions and repeat the regression process. Once the algorithm converges after a number of directional sweeps is performed, the separation rank $r$ is increased if the residual norm $\Vert u-u_r\Vert_{D}$ is above a prescribed accuracy $\epsilon$. The ALS approach can be thought of as a non-linear generalization of the block Gauss-Seidel method as we solve for one set of variables at a time while fixing all the others. We note that such a multi-linear approximation is possible, particularly, due to the separated form of $u_r$ with respect to $y_k$'s.\\

\noindent {\it Linear least-squares regression along $y_k$.} For a given $k\in\{1,\dots,d\}$ and some $r\ge 1$, assume that scalars $\{s_l\}$ and functions $\{u_i^{l}(y_i)\}$, $i\ne k$, $l=1,\dots,r$, are given and fixed. We then seek the unknown univariate functions $\{u_k^{l}(y_k)\}$, $l=1,\dots,r$, from the restriction of (\ref{eq:low-rank_regression}) to the direction $k$. That is, 
\begin{eqnarray}
\label{eq:1d_opt}
\{u_k^{l}(y_k)\} 
&=& \arg\min_{\{\hat{u}_k^{l}(y_k)\} }\Vert u-\hat{u}_r \Vert_{D}^2\\
&=&\arg\min_{\{\hat{u}_k^{l}(y_k)\} }\frac{1}{N}\sum_{j=1}^N \left(u(\bm{y}^{(j)})-\sum_{l=1}^r \hat{u}_k^l (y_k^{(j)}) s_l \prod_{i\ne k} u_i^l (y_i^{(j)} )\right)^2,\nonumber 
\end{eqnarray}
$l=1,\dots,r$, which is in the form of a standard one-dimensional regression of scattered data. 

A discrete formulation of (\ref{eq:1d_opt}) can then be obtained by expanding each (unknown) univariate function $u_k^{l}(y_k)$ (or $\hat{u}_k^{l}(y_k)$) into some finite-dimensional basis of $L_{2}(\Gamma,\mathcal{P})$. In the present study, we choose spectral polynomials orthogonal with respect to $\rho(y_k)$ -- the probability density function of $y_k$ -- for this purpose. We note that implicit in such a selection is the assumption that each factor $u_k^{l}(y_k)$ depends smoothly on $y_k$.\\[-.3cm]


\noindent {\it Spectral polynomial expansion of univariate factors $u_k^l(y_k)$.} Let $\{\psi_{\alpha_k}(y_k)\}$ be the set of spectral polynomials of degree $\alpha_k\le M\in \mathbb{N}_0$ associated with $\rho(y_k)$ as introduced in Section \ref{sec:Separated Representation}. We then approximate
\begin{equation}
\label{eq:factor_expansion}
u_k^l(y_k)\approx \sum_{\alpha_k=0}^{M}c_{\alpha_k}^{l}\psi_{\alpha_k}(y_k), \qquad l=1,\dots,r,
\end{equation}
in which, for the interest of simplicity, we assumed the maximum degree of polynomial expansion, $M$, is the same for all separation terms and in all directions $k$. However, this assumption may be relaxed with no technical difficulty. In fact, for situations where the relative importance of the random inputs $y_k$ is available, an anisotropic selection of $M$ may considerably improve the convergence of the approach.

Given the finite-dimensional approximation of $u_k^l(y_k)$ in (\ref{eq:factor_expansion}), the problem (\ref{eq:1d_opt}) reduces to the discrete least-squares optimization 
\begin{eqnarray}
\label{eq:1d_opt_dicrete}
\bm{c}_{k}
&=& \arg\min_{\hat{\bm c}_{k}}\Vert u-\hat{u}_r \Vert_{D}^2\\
&=&\arg\min_{\hat{\bm c}_{k}}\frac{1}{N}\sum_{j=1}^N \left(u(\bm{y}^{(j)})-\sum_{l=1}^r \left(\sum_{\alpha_k=0}^{M}\hat{c}_{\alpha_k}^{l}\psi_{\alpha_k}(y_k^{(j)})\right) s_l \prod_{i\ne k} u_i^l (y_i^{(j)} )\right)^2\nonumber 
\end{eqnarray}
to compute the expansion coefficients $\bm{c}_{k} := (c_0^1,\cdots,c_M^{1},\cdots,c_0^r,\cdots,c_M^{r})^{T}\in\mathbb{R}^{r(M+1)}$ along direction $k$. Setting the derivative of $\Vert u-\hat{u}_r \Vert_{D}^2$ with respect to $\bm{c}_{k}$ to zero, we arrive at the normal equation
\begin{equation}
\label{eq:normal}
\bm{A}_k^T \bm{A}_k \bm{c}_k= \bm{A}_k^T \bm{u},
\end{equation}
to solve for $\bm{c}_{k}$. Here, the matrix $\bm{A}_k \in \mathbb{R}^{N\times r (M+1)}$ has a column block-structure $\bm{A}_k = \left[\bm{A}_k^{1}\ \dots\ \bm{A}_k^{r}\right]$, where each column-block $\bm{A}_k^{l}\in \mathbb{R}^{N\times (M +1)}$, $l=1,\dots,r$, is given by
\begin{equation}
\label{eq:mat_A} 
\bm{A}_k^{l}(j,\alpha_k+1) = s_l \psi_{\alpha_k}(y_k^{(j)}) \prod_{i\ne k} u_i^l (y_i^{(j)}),
\end{equation}
for $j=1,\dots,N$ and $\alpha_k=0,\dots,M$. Moreover, the data vector $\bm u\in\mathbb{R}^{N}$ contains realizations of $u(\bm y)$, i.e., $\bm{u}(j)=u(\bm{y}^{(j)})$, $j=1,\dots,N$. 

Having $\bm c_k$ computed from (\ref{eq:normal}), we then update each $s_l$ such that $s_l\leftarrow s_l\Vert u_k^{l}\Vert_D$ and accordingly set $\bm c_k \leftarrow \bm c_k / \Vert u_k^{l}\Vert_D$. We do so to conveniently compare the relative contribution of each separation term. Additionally, as described in \cite{Beylkin02,Beylkin05}, such a normalization of $\bm{c}_k$ may be used to control the loss of significant digits in the evaluation of the separated representation. \\[-.3cm]

\noindent {\it The alternation to $y_{k+1}$ and separation rank increase.} Once the factors $\{u_k^l(y_k)\}$, $l=1,\dots,r$, are computed based on $\bm c_k$, as discussed above, the algorithm alternates to direction $k+1$ in order to compute $\{u_{k+1}^l(y_{k+1})\}$, $l=1,\dots,r$, using recent estimates of $\{u_i^l(y_i)\}_{i=1}^k$, $l=1,\dots,r$. The process will continue until all factors $\{u_k^l(y_k)\}_{k=1}^d$, $l=1,\dots,r$, are updated. In practice, we initialize the the factors $\{u_k^l(y_k)\}_{k=1}^d$, $l=1,\dots,r$, randomly (or according to some other guess); therefore, the sweeps through directions $k$ must be performed multiple times until the residual norm $\Vert u-u_r \Vert_{D}$ does not decrease further. Notice that, by the ALS construction, $\Vert u-u_r \Vert_{D}$ monotonically decreases or remains constant throughout the iterations. For a given separation rank $r$, if the residual norm $\Vert u-u_r \Vert_{D}$ at the end of the ALS updates is not smaller than a target accuracy $\epsilon$, then $r$ is increased to $r+1$ and the algorithm is repeated. The overall non-intrusive ALS procedure is summarized in Algorithm \ref{Algorithm:ALS}.\\

\begin{algorithm}[H]
\SetKwData{Left}{left}
\SetKwData{This}{this}
\SetKwData{Up}{up}
\SetKwFunction{Union}{Union}
\SetKwFunction{FindCompress}{FindCompress}
\SetKwInOut{Input}{ Input}
\SetKwInOut{Output}{ Output}
$\bullet$\Input{Data set $D=\left\lbrace\left( \bm{y}^{(j)}; u(\bm{y}^{(j)})\right)\right\rbrace_{j=1}^N$, polynomial orders $M$, and accuracy $\epsilon$}
\BlankLine
$\bullet$\Output{$r$, $\{c_{\alpha_k}^l\}_{k=1}^d$, and $s_l$ for $\alpha_k=0,\dots,M$ and $l=1,\dots,r$}
\BlankLine
$\bullet$ Set $r = 1$; (randomly) initialize $\{c_{\alpha_k}^r\}$ (or equivalently $\{u_k^r(y_k)\}$)
\BlankLine
\While{$\Vert u-u_r\Vert_D > \epsilon$}{
\BlankLine
	\While{$\Vert u-u_r\Vert_D$ decreases much}{
		\BlankLine
	 	\For{$k\leftarrow 1$ \KwTo $d$}{
		\BlankLine
		$\bullet$ Fix $\lbrace c_{\alpha_i}^l \rbrace$, $i\ne k$, and solve for $\lbrace c_{\alpha_k}^l\rbrace$ using (\ref{eq:normal})\\
		\BlankLine
		$\bullet$ Update $s_l\leftarrow s_l\  \Vert u_k^l\Vert_D$ and $c_{\alpha_k}^{l}\leftarrow c_{\alpha_k}^{l}/ \Vert u_k^l\Vert_D$
		\BlankLine
		}		
       \BlankLine
	}
	\BlankLine
	$\bullet$ Set $r = r+1$; (randomly) initialize $\{c_{\alpha_k}^{r}\}$ (or equivalently $\{u_k^r(y_k)\}$) and, thus, $s_r$ for $k=1,\dots,d$
	\BlankLine
}
\label{Algorithm:ALS}
\caption{Summary of the ALS method for the non-intrusive construction of separated representations.}
\end{algorithm}

\begin{rem}
While in the preceding descriptions and in Algorithm \ref{Algorithm:ALS} we require $\epsilon$ as an input to our construction, in practice $\epsilon$ may not be known {\it a priori} and a rough estimate of $\epsilon$ may lead to under-fitting or over-fitting. We address this issue in Section \ref{sec:overfitting}, where we instead select an ``optimal'' value of $r$ (and $M$) based on an error indicator.  Therefore, no stopping criterion $\epsilon$ (per se)  is required. 
\end{rem}

\begin{rem}
For each $k$, as the coefficients $\{c_{\alpha_k}^{l}\}$ corresponding to different values of $l$ are coupled in the linear systems (\ref{eq:normal}), the factors $\{u_k^l(y_k)\}$, $l=1,\dots,r$, may change upon the introduction of the new separation term (to increase $r$). Alternatively, following the work in \cite{Ammar06,LeBris09}, a greedy approach may be applied to perform updates only on the new separation term while freezing all the previous ones. The investigation of this approach is, however, beyond the scope of the present study.   
\end{rem}

\noindent{\it Notation Alert.} For a simpler presentation, unless required, we hereafter eliminate the subscripts $k$ from the linear systems in (\ref{eq:normal}) and their extensions. 

\subsection{Solution statistics}
\label{sec:rs}

Given the spectral coefficients $\{c_{\alpha_k}^{l}\}$, one can estimate the statistics of $u$ by integrating $u_r$ analytically, e.g., for the mean and variance, numerically via quadrature integration, e.g., for the kurtosis, or via random sampling, e.g., for the probability density function. Specifically, the integral-form statistics, e.g., moments of $u$, can be obtained as a sequence of one-dimensional integrals, thanks to the separated form of $u_r$. For instance, the estimate of the mean and second moment of $u$ is given by 
\begin{eqnarray}
\label{eqn:mean}
\mathbb{E}\left[u_r\right]&=&\int_{\Gamma^d} u_r(\bm y)\rho(\bm y)d\bm y = \sum_{l=1}^{r}s_{l}\prod_{k=1}^{d}\left(\int_{\Gamma}u_{k}^{l}(y_{k})\rho_{k}(y_k)dy_{k}\right)\nonumber\\
&=& \sum_{l=1}^{r}s_{l}\prod_{k=1}^{d} c_{0}^{l}
\end{eqnarray}
and
\begin{eqnarray}
\label{eqn:2nd}
\mathbb{E}\left[u_r^2\right]&=&\int_{\Gamma^d} u_r^2(\bm y)\rho(\bm y)d\bm y = \sum_{l=1}^{r}\sum_{l'=1}^{r}s_{l}s_{l'}\prod_{k=1}^{d}\left(\int_{\Gamma}u_{k}^{l}(y_{k})u_{k}^{l'}(y_{k})\rho_{k}(y_k)dy_{k}\right)\nonumber\\
&=& \sum_{l=1}^{r}\sum_{l'=1}^{r}s_{l}s_{l'}\prod_{k=1}^{d}\left(\sum_{\alpha_k = 0}^M c_{\alpha_k}^{l}c_{\alpha_k}^{l'}\right),
\end{eqnarray}
respectively, where we used the orthonormality of the spectral basis $\{\psi_{\alpha_k}(y_k)\}$. In general, the $m$-th moments of $u_r$ can be similarly computed based on the expectations of the product of $m$ spectral polynomials $\{\psi_{\alpha_k}(y_k)\}$, which may be computed exactly using Gaussian quadrature integration.  

Alternatively, one may randomly draw independent realizations $\{u_r(\bm y^{(j)})\}$ of $u_r$ to generate Monte Carlo estimates of the statistics of $u_r$.

\subsection{Computational cost of ALS and dependence of $N$ on $d$}
\label{sec:cost}

Here we elaborate on the computational complexity of a full sweep of the ALS and quantify its dependence on $d$ and $N$. For this purpose, we consider the asymptotic (but relevant) case of $N\gg rM$. Each factor $u_k^l(y_k)$ can be evaluated with complexity $\mathcal{O}(M)$ for a given $y_k^{(j)}$. Therefore, the matrices $\bm{A}$ in (\ref{eq:normal}) are generated with a cost of $\mathcal{O}(rdMN)$. Each normal equation (\ref{eq:normal}) may be solved using the Cholesky or LU decomposition of $\bm{A}^T \bm{A}$ with an asymptotic complexity $\mathcal{O}(r^2M^2N)$ that includes forming the normal matrix $\bm{A}^T \bm{A}$ and the right-hand-side vector $\bm{A}^T \bm{u}$ as well as computing the solution $\bm{c}$. We here assumed that $r^3M^3\ll r^2M^2 N$.

Following \cite{Beylkin08}, to alternate from direction $k$ to $k+1$, the $\bm{A}_{k+1}$ matrices may be updated from $\bm{A}_k$ by multiplying products $\prod_{i\ne k} u_i^l (y_i^{(j)})$ in (\ref{eq:mat_A}) by $u_k ^l(y_k^{(j)})/u_{k+1}^l(y_{k+1}^{(j)})$. This requires $\mathcal{O}(rdMN)$ cost for a full sweep instead of $\mathcal{O}(rd^2MN)$ if $\bm{A}_{k+1}$ was set directly. Adding up the complexities of forming and solving the normal equations (\ref{eq:normal}) for all directions $k=1,\dots,d$, we arrive at 
\begin{equation}
\label{eq:full_complexity}
\mathcal{O}(r^2dM^2N)
\end{equation}
for the cost of a full sweep of the ALS iteration. Given that the number $N$ of solution realizations generally depends on $d$, the estimate in (\ref{eq:full_complexity}) is in fact super-linear in $d$. We next discuss the dependence of $N$ on $d$. \\[-.3cm]

\noindent {\it Dependence of $N$ on $d$.} As opposed to the PC expansions (\ref{eq:pce_multi}) where the number of unknown coefficients $s_{\bm\alpha}$ grows exponentially with $d$, the total number of unknown coefficients in the separated representation (\ref{eq:separated_intro}) is $rd(M+1)$, which depends linearly on $d$, assuming that the separation rank $r$ is independent of $d$. We, therefore, expect a linear dependence of $N$ on $d$, i.e.,  
\begin{equation}
\label{eqn:order_samples}
N\sim \mathcal{O}(rdM),
\end{equation}
for a successful separated approximation. Plugging (\ref{eqn:order_samples}) in (\ref{eq:full_complexity}) leads to an overall computational complexity
\begin{equation}
\label{eq:full_complexity_with_samples}
\mathcal{O}(r^3d^2M^3) \nonumber
\end{equation}
for a full sweep of the ALS iteration, which is quadratic in $d$. We highlight that for situations where evaluating the solution of the forward model is significantly expensive this overall cost is reasonable.  

\begin{rem}
The estimate provided in (\ref{eqn:order_samples}) is based on the rule-of thumb that a successful regression problem requires a number of data points that is multiple of the number of unknowns. However, a more precise estimate of this dependence may be needed. In particular, the number of unknown coefficients in the one-dimensional regression problems (\ref{eq:1d_opt_dicrete}) is $r(M+1)$, hence independent of $d$. Moreover, the same samples used to compute the coefficients $\bm{c}_{k}$ in (\ref{eq:normal}) are also utilized to compute $\bm{c}_{k+1}$, and so on. Therefore, the linear dependence of $N$ on $d$ may be relaxed. On the other hand, in the absence of the regularization approach of Section \ref{sec:Tikhonov}, $N$ may need to grow like $\mathcal{O}(r^\zeta M^\zeta)$, $\zeta > 1$, to keep the condition number of the matrices $\bm A_{k}$ small; therefore, the linear dependence of $N$ on $r$ and $M$ in (\ref{eqn:order_samples}) may be optimistic. 
\end{rem}


%
\section{Preventing over-fitting: regularization and selection of $r$ and $M$}
\label{sec:overfitting}

Similar to other regression methods, the non-intrusive separated representation (\ref{eq:separated_intro}) may over-fit the data, especially when the number of solution samples, $N$, is small. By the ALS construction, the residual norm $\Vert u - u_r\Vert_D$ decreases or remains unchanged throughout the one-dimensional updates as well as when the separation rank $r$ is increased. As $u$ is not exactly separable or may be noisy, excessive reduction of $\Vert u - u_r\Vert_D$ results in over-fitting which may be accompanied by poor approximations of $u$ where data is not available. Additionally, an unnecessarily large degree $M$ of the spectral approximation of factors $\{u_k^l(y_k)\}$ may also lead to over-fitting in the one-dimensional regression problems (\ref{eq:1d_opt_dicrete}). Numerically, these two over-fitting scenarios give rise to ill-conditioned matrices $\bm{A}$ in (\ref{eq:normal}), hence, inaccurate solutions $\bm{c}$ and subsequent ALS updates. 

The naive approach to reduce, or even prevent, over-fitting is to choose {\it small} values for $r$ and $M$. However, this may lead to inaccurate (under-fitted) solution $u_r$. A more formal strategy is based on the concept of regularization, where additional smoothness constraints are enforced on $u_r$ while allowing for moderate values of $r$ and $M$.   

In the present study, we examine a Tikhonov regularization approach to reduce the over-fitting for given choices of $r$ and $M$. Subsequently, in Section \ref{sec:Perturbation_based_error}, we develop a perturbation-based error indicator based on the regularized version of problem (\ref{eq:normal}) in order to identify the values of $r$ and $M$ leading to stable and, at the same time, accurate solutions.

\subsection{Tikhonov regularization}
\label{sec:Tikhonov}

One classical way to regularize the solution of regression -- or more generally inverse -- problems is through the use of {\it Tikhonov} regularization, where a smoothness promoting penalty term is added to the regression cost function, e.g., see \cite{Hansen98b}. While the non-intrusive construction of the separated representation is originally a non-linear problem, due to the multi-linear attribute of the ALS, the Tikhonov regularization can be naturally applied to the linear regression problems (\ref{eq:1d_opt_dicrete}). Specifically, consider the matrix formulation of (\ref{eq:1d_opt_dicrete}), i.e., 
\begin{equation}
\label{eq:1d_opt_dicrete_matrix}
\bm{c} = \arg\min_{\hat{\bm c}}\; \frac{1}{N}\left\Vert \bm{A} \hat{\bm{c}}- \bm{u} \right\Vert_2^2.
\end{equation}

In Tikhonov regularization, a smoothness penalty of the form $\Vert \bm{L}{\bm c}\Vert_2^2$,  $\bm{L}\in\mathbb{R}^{r(M+1)\times r(M+1)}$, is enforced on the solution $\bm c$ using the method of Lagrange multipliers, i.e., 
\begin{equation}
\label{eq:Tikhonov regularization}
\bm{c}_\lambda = \arg\min_{\hat{\bm c}}\; \frac{1}{N}\left\Vert \bm{A} \hat{\bm{c}}- \bm{u} \right\Vert_2^2+\lambda^2 \Vert \bm{L}\hat{\bm c}\Vert_2^2.
\end{equation}

Here, $\bm{L}$ and $\lambda\in\mathbb{R}_{\ge 0}$ are called the {\it Tikhonov matrix} and {\it regularization parameter}, respectively. The regularization parameter $\lambda$ balances the relative contributions of the mismatch function $\frac{1}{N}\left\Vert \bm{A} {\bm{c}}- \bm{u} \right\Vert_2^2$ and the penalty term $\Vert \bm{L}{\bm c}\Vert_2^2$.

It is straightforward to show that the solution to the Tikhonov-regularized problem (\ref{eq:Tikhonov regularization}) can be computed from the modified normal equation
\begin{equation}
\label{eq:normal_tikh}
\left(\bm{A}^T \bm{A}+\lambda^2 \bm{L}^T \bm{L}\right) \bm{c}_\lambda= \bm{A}^T \bm{u}.
\end{equation}

The effectiveness of the Tikhonov regularization depends on the choices of $\bm L$ and $\lambda$. We next describe the approaches we followed in this study for the selection of these quantities. 

\subsubsection{Choice of Tikhonov matrix $\bm L$}
\label{sec:tikh_matrix}

The standard form of the Tikhonov matrix is when $\bm L=\bm I$, the identity matrix. For linear regression problems with orthonormal  basis, $\Vert \bm{I}{\bm c}\Vert_2^2$ is essentially the second moment of the approximate solution. Therefore, large deviations in regions where data is highly sparse or not available will be penalized in the reconstruction. However, as demonstrated in the numerical example of Section \ref{sec:Hydrogen Oxidation Problem}, the choice of $\bm L=\bm I$, or more precisely $\bm{L} = \bm{I}_{M+1}\otimes diag(s_1,\dots,s_r)$ to account for the normalization of $\bm c$,  may lead to an inadequate regularization of (\ref{eq:1d_opt_dicrete_matrix}). Here $\bm{I}_{M+1}$ and $\otimes$ denote the identity matrix of size $M+1$ and the Kronecker product, respectively. The reason for this inadequacy is because no orthogonality condition among the separation terms is enforced and $\Vert (\bm{I}_{M+1}\otimes diag(s_1,\dots,s_r)){\bm c}_k\Vert_2^2 = \sum_{l=1}^{r}s_l^2\mathbb{E}[(u_k^l)^2]\ne \mathbb{E}[u_r^2]$ ignores the contributions of factors along directions $i\ne k$. Therefore, in the present study, we derive the regularization matrix $\bm L$ such that
\begin{equation}
\label{eq:2nd_moment_reg}
\Vert \bm{L} \bm{c} \Vert_2^2 = \mathbb{E}[u_r^2], 
\end{equation}
where, here, $u_r$ refers to the separated approximation of $u$ at the iteration in which the regularization is performed. Following (\ref{eqn:2nd}), it is straightforward that $\mathbb{E}[u_r^2] = \bm c^{T}\bm{B}\bm {c}$, where the $(l,l')$-th block of the symmetric, positive-definite matrix $\bm B\in\mathbb{R}^{r(M+1)\times r(M+1)}$ is given by
\begin{equation} 
\label{eq:block of B matrix}
\bm{B}{(l,l')}= s_l s_{l'} \prod_{i\ne k} \left(  \sum_{\alpha_i=0}^M c_{\alpha_i}^{l} c_{\alpha_i}^{l'} \right) \bm{I}_{M+1}, \quad l,l'=1,\dots,r.\nonumber
\end{equation} 

Using  (\ref{eq:2nd_moment_reg}) and assuming that $\mathbb{E}[u_r^2]>0$, the Tikhonov matrix $\bm L$ can be set from the Cholesky decomposition of $\bm B$, i.e., 
\begin{equation} 
\label{eq:our_L}
\bm B = \bm L^T\bm L.
\end{equation} 
%

%
%

While we here choose $\bm L$ according to the condition (\ref{eq:2nd_moment_reg}), other constructions, for instance, based on the gradient of $u_r$ with respect to $\bm y$ may also be effective in promoting smoothness, see, e.g., \cite{dAvezac11}. 

\begin{rem}
\label{rem:L_pos}
Assuming that $\mathbb{E}[u_r^2]>0$, the Tikhonov matrix $\bm L$ in (\ref{eq:our_L}) is invertible. This property will be exploited in the analysis of Section \ref{sec:Perturbation_based_error}. 
\end{rem}

\subsubsection{Selection of regularization parameter $\lambda$: Generalized Cross Validation (GCV)}
\label{sec:GCV}

Selecting a {\it suitable} value for the regularization parameter $\lambda$ is essential to the success of the Tikhonov regularization. A substantially underestimated $\lambda$ results in non-smooth solutions due to the over-fitting. On the other hand, a significantly overestimated $\lambda$ leads to highly biased and smooth solutions. In both cases, the approximation of $u_r$ may be inaccurate. Several statistical methods including Unbiased Predictive Risk Estimator, \cite{Mallows73,Craven79}, Morozov's Discrepancy Principle, \cite{Morozov84}, L-curve, \cite{Hansen92,Hansen93}, and Generalized Cross Validation (GCV), \cite{Wahba77,Golub79}, have been proposed to estimate $\lambda$. Assuming the samples $\bm u$ are noise-free, the first two approaches require an accurate knowledge of the second moment of the mismatch $u-u_r$. Such information is not generally available, especially that the second moment of $u-u_r$ depends on $r$ and $M$ and varies throughout the ALS iterations. The last two methods above, however, do not require such information and may be applied. 

In the present study, we employ the GCV technique to select $\lambda$. GCV is a modification of the classical {\it leave-one-out} Cross Validation (CV) where, for a fixed $\lambda$, the regression is performed using $N-1$ sample points and the resulting approximation is tested -- or {\it validated} -- against the remaining one. Such calibration and validation are performed on the $N$ possible leave-one-out scenarios and the validation error is averaged. The process is then repeated for several values of $\lambda$ and the CV selection of $\lambda$ is the one associated with the smallest average validation error. Specifically, let
\begin{equation}
\label{eq:gcv_orig}
\bm{c}_{\lambda}^{[j]} = \arg\min_{\hat{\bm c}}\; \frac{1}{N-1}\left\Vert \bm{A}^{[j]} \hat{\bm{c}}- \bm{u}^{[j]} \right\Vert_2^2+\lambda^2 \Vert \bm{L}\hat{\bm c}\Vert_2^2
\end{equation}
be the solution of the Tikhonov-regularized problem (\ref{eq:1d_opt_dicrete_matrix}) for some value of $\lambda$ and when the realization $\bm{u}(j)$ is excluded from the dataset $D$. Then the CV choice of $\lambda$ is given by
\begin{equation}
\label{eq:gcv_fun_orig}
\lambda = \arg\min_{\hat\lambda}\; CV(\hat\lambda) = \frac{1}{N} \sum_{j=1}^{N} \left((\bm{A}\bm{c}_{\hat\lambda}^{[j]})(j) - \bm{u}(j)\right)^2,  
\end{equation}
which is equivalent to
\begin{equation}
\label{eq:GCV_fun_mod}
\lambda = \arg\min_{\hat\lambda}\; CV(\hat\lambda)=  \frac{1}{N} \sum_{j=1}^N\left(\frac{(\bm{A}\bm{c})(j) - \bm{u}(j)}{1-\bm{H}_{\hat\lambda}(j,j)} \right)^2,
\end{equation}
see, e.g., \cite{Wahba77}. Here, the \textit{hat matrix} $\bm{H}_{\lambda}=\bm{A}(\bm{A}^T \bm{A}+\lambda^2 \bm{L}^T \bm{L} )^{-1} \bm{A}^T$ maps the vector of the realizations $\bm u$  to their predicted values based on the separated representation. Notice that the evaluation of the $CV$ function in (\ref{eq:GCV_fun_mod}) is easier than that of the (\ref{eq:gcv_fun_orig}) as, for each $\lambda$, only one calibration is performed in (\ref{eq:GCV_fun_mod}). The diagonal entries $\bm{H}_\lambda(j,j)$ depend on the ordering of the rows in $\bm{A}$, hence the ordering of realizations $\{\bm y ^{(j)}\}$. To address this drawback, Generalized Cross Validation (GCV) was introduced in \cite{Wahba77,Golub79}, where $\bm{H}_\lambda(j,j)$ is approximated by the average of the trace of $\bm H_\lambda$, that is $\bm{H}_{\lambda}(j,j)\approx \frac{1}{N}tr(\bm H_\lambda)$. Therefore, in GCV the selection of $\lambda$ is given by
\begin{equation}
\label{eq:GCV_approx}
\lambda = \arg\min_{\hat\lambda}\;  GCV(\hat\lambda)=\frac{ N\left\Vert \bm{A} \bm{c}- \bm{u} \right\Vert_2^2}{ \left( N-tr(\bm{H}_{\hat\lambda}) \right)^2 }.
\end{equation}

\begin{rem}
We note that the range of $\lambda$ for which the GCV (or CV) function is evaluated to find the minimum can be set based on the (generalized) singular values of $\bm A$. For further information on the selection of this range as well as the details on how GCV is efficiently evaluated in practice, we refer to \cite{Hansen92,Vogel02, Hansen10}.  
\end{rem}

\subsection{Perturbation-based error indicator for selection of $r$ and $M$}
\label{sec:Perturbation_based_error}

By construction, the ALS approach monotonically decreases the residual norm $\Vert u - u_r\Vert_D$ throughout the iterations with both fixed separation rank $r$ and when $r$ is increased. However, due to the over-fitting issue discussed at the beginning of Section \ref{sec:overfitting}, a large $r$ may result in a solution $u_r$ that is far from $u$ where samples are not available. The regularization approach of Section \ref{sec:Tikhonov} mitigates this issue to a great extent, however, it does not provide a stopping criterion for the separation rank increase. Similarly, given a fixed number $N$ of solution realizations, there is the need to identify a value for $M$, the degree of spectral basis, which does not lead to a highly under-fitted or over-fitted approximation. 

To this end, we propose the selection of $r$ and $M$ using a {\it perturbation} bound on the sensitivity of the regularized solution of (\ref{eq:normal_tikh}) to the mismatch between $u$ and $u_r$ in (\ref{eq:separated_intro}). In particular, for a given $M$, the separation rank $r$ is increased until the sensitivity of the solution to (\ref{eq:normal_tikh}) across all directions $k$ increases. This process is repeated for different values of $M$ and the pair $(r,M)$ is chosen such that the sensitivities of the associated regularized solutions (largest across directions $k=1,\dots,d$) is the minimum among other pairs $(r,M)$. 
More precisely, let 
\begin{equation}
\label{eqn:error_samp}
\varepsilon({\bm{y}^{(j)}}) = u(\bm y^{(j)}) - u_r(\bm y^{(j)}),\quad j=1,\dots,N,
\end{equation}
be the realization of the separated representation error $\varepsilon$ in (\ref{eq:separated_intro}) where $u_r$ is the separated representation of $u$ obtained with a fixed separation rank $r$, spectral polynomial degree $M$, but with a sufficiently large number of samples $N_{e}\gg N$. For a moment assume that the vector of separated representation errors $\bm\varepsilon\in\mathbb{R}^{N}$ with $\bm{\varepsilon}(j)=\varepsilon({\bm{y}^{(j)}})$, $j=1,\dots,N$, is known. Then the sensitivity of the Tikhonov solution $\bm c_\lambda$ of (\ref{eq:normal_tikh}) under this error, or perturbation of the realizations of $u_r$, may be bounded as follows.

\begin{prop}[Perturbation bound]
Let $\bm c_{\lambda}$ and $\tilde{\bm c}_{\lambda}$ be solutions to the problem (\ref{eq:Tikhonov regularization}) associated with data vectors $\bm u$ and $\bm u - \bm{\varepsilon}$, respectively, where the perturbation vector $\bm\varepsilon\in\mathbb{R}^N$ with $\bm{\varepsilon}(j)=\varepsilon({\bm{y}^{(j)}})$, $j=1,\dots,N$, is given in (\ref{eqn:error_samp}). Then,
\begin{equation}
\label{eqn:perturb_bound}
\frac{\Vert \bm{c}_{\lambda}-\tilde{\bm{c}}_{\lambda}\Vert_2}{\Vert \bm{c}_{\lambda}\Vert_2}\leq \frac{\lambda^{-1} \Vert \bm{L}^{-1} \Vert_2\ \Vert\bm\varepsilon\Vert_2}{\Vert \bm{c}_{\lambda}\Vert_2}.
\end{equation}
\end{prop}

\noindent{\it Proof.} The proof is a straightforward simplification of Theorem $5.5.1$ in \cite{Hansen98b}. We only remind that, according to Remark \ref{rem:L_pos}, $\bm L^{-1}$ exists. $\square$  \\[-.3cm]

Notice that the bound in (\ref{eqn:perturb_bound}) depends on $\Vert\bm\varepsilon\Vert_2$ which is not readily available in practice. However, under some assumptions, the expected value of  $\Vert\bm\varepsilon\Vert_2^2$ may be estimated. In particular, we assume that $\varepsilon({\bm{y}^{(j)}})$, $j=1,\dots, N$, can be approximated by realizations of some uncorrelated random variables with zero mean and variance $\sigma^2$. Then, following \cite{Green94},  $\sigma^2$ may be estimated by
\begin{equation}\label{eq:variance estimator}
\widehat{\sigma}_\lambda^2\ = \frac{\left\Vert \bm{A} \bm{c}_\lambda - \bm{u} \right\Vert_2^2}{ \left( N-tr(\bm{H}_\lambda) \right) }.
\end{equation}

Consequently, we have
\begin{equation}
\label{eqn:mean_est}
\mathbb{E}\left[\Vert \bm\varepsilon\Vert_2^2\right] = N\widehat{\sigma}_\lambda^2.
\end{equation}

We then define an {\it error indicator} of the form
\begin{equation}
\label{eqn:ei}
EI = \frac{N^{\frac{1}{2}}\lambda^{-1} \Vert \bm{L}^{-1} \Vert_2\ \widehat{\sigma}_\lambda}{\Vert \bm{c}_{\lambda}\Vert_2},
\end{equation}
which is similar to the perturbation bound in (\ref{eqn:perturb_bound}), except that we replaced $\Vert\bm\varepsilon\Vert_2^2$ with its expected value $N\widehat{\sigma}_\lambda^2$. 

In the numerical examples of Section \ref{sec:num_exp}, we use the error indicator $EI$ in (\ref{eqn:perturb_bound}) to estimate a pair $(r,M)$ that does not lead to over-fitting or under-fitting. Specifically, for a fixed $M$, we monitor $EI$ throughout the last $d$ iterations of the ALS updates immediately before $r$ is increased. We then take $EI_{\max}^{r,M}$ as the largest $EI$ among these $d$ indicators. Our selection of the separation rank $r$ and polynomial degree $M$, hence the stopping criterion of the separated representation, is the pair $(r,M)$ corresponding to the smallest $EI_{\max}^{r,M}$. Notice that the $EI$ in (\ref{eqn:ei}) implicitly depends on $r$ through $\lambda^{-1}$, $\Vert \bm{L}^{-1} \Vert_2$, $\widehat{\sigma}_\lambda$, and $\Vert \bm{c}_{\lambda}\Vert_2$. Roughly speaking, when $r$ is too small or unnecessarily large, $EI$ is large due to $\widehat{\sigma}_\lambda$ or $\Vert \bm L^{-1}\Vert_2$ being large, respectively. For values of $r$ close to the {\it optimal} separation rank (e.g., one resulting in minimum mean-squares error), however, these latter quantities are moderate. This is why the pair $(r,M)$ obtained via $EI$ may avoid under-fitting and over-fitting in the separated approximation.

\begin{rem}
We call the quantity in (\ref{eqn:ei}) an error {\it indicator} as opposed to an error {\it estimator} as (\ref{eqn:ei}) does not account for the bias introduced in the estimation of $\bm c$ due to the Tikhonov regularization. Additionally, due to the iterative construction of ALS, the matrices $\bm A$ are affected by the errors in the estimation of $\bm c$ coefficients from the previous iterations. While the effect of such errors may be modeled in the form of perturbations of $\bm A$, we ignore this effect in the present study. Even so, as we shall see in numerical results of Section \ref{sec:num_exp}, the error indicator in (\ref{eqn:ei}) successfully identifies the pair $(r,M)$ that avoids over-fitting and at the same time leads to accurate separated representation. 
\end{rem}

We next investigate the performance of the non-intrusive separated representation together with the proposed regularization strategy on a manufactured function and two ODE models subject to high-dimensional random inputs.


%

%
\section{Numerical Experiments}
\label{sec:num_exp}
\subsection{A manufactured function}
\label{sec:manufactured problem}

Here we consider the separated representation of the function 
\begin{equation}
\label{eq:manufactured function}
u(y_1,\dots,y_{10})= s_0+ s_1\psi_3(y_1)\psi_3(y_2) + s_2\psi_2(y_3) + s_3 \psi_2(y_8) + s_4\psi_3(y_9) + \varepsilon,
\end{equation}
where $\{y_k\}_{k=1}^{10}$ are i.i.d standard Gaussian random variables and $\psi_i(\cdot)$ is the Hermite polynomial of degree $i$, normalized to have variance one. The noise $\varepsilon$ is a zero mean Gaussian random variable with standard deviation $0.0005$ and is independent from $\{y_k\}_{k=1}^{10}$. The coefficients $\{s_i\}_{i=1}^{4}$ are set to $s_0=0.55$, $s_1=\sqrt{2}/2$, $s_2=-\sqrt{2}/4$, $s_3=-\sqrt{2}/4$, and $s_4=-1/10$. This setting leads to $\mathbb{E}[u]=0.55$ and $\mathrm{Var}[u] = 0.76$ for the noise-free $u$.

The purpose of this test case is to verify the selection of the separation rank $r$ and spectral polynomial degree $M$ via the error indicator of Section \ref{sec:Perturbation_based_error}.  In particular, we expect that $r\le 5$ as the the separated representation compresses the noise-free $u$ with a separation rank that cannot exceed the number of orthogonal terms in (\ref{eq:manufactured function}), i.e., 5. This is due to the construction of separated representations where no orthogonality condition among the separated terms is assumed. Additionally, an accurate separated approximation of $u$ requires an order $M\ge 3$ for the spectral expansion of the factors in (\ref{eq:separated_intro}). 

\begin{figure}
    \centering
    \begin{tabular}{cc}
            \hspace{-0.5cm}    
      \includegraphics[width=2.8in]{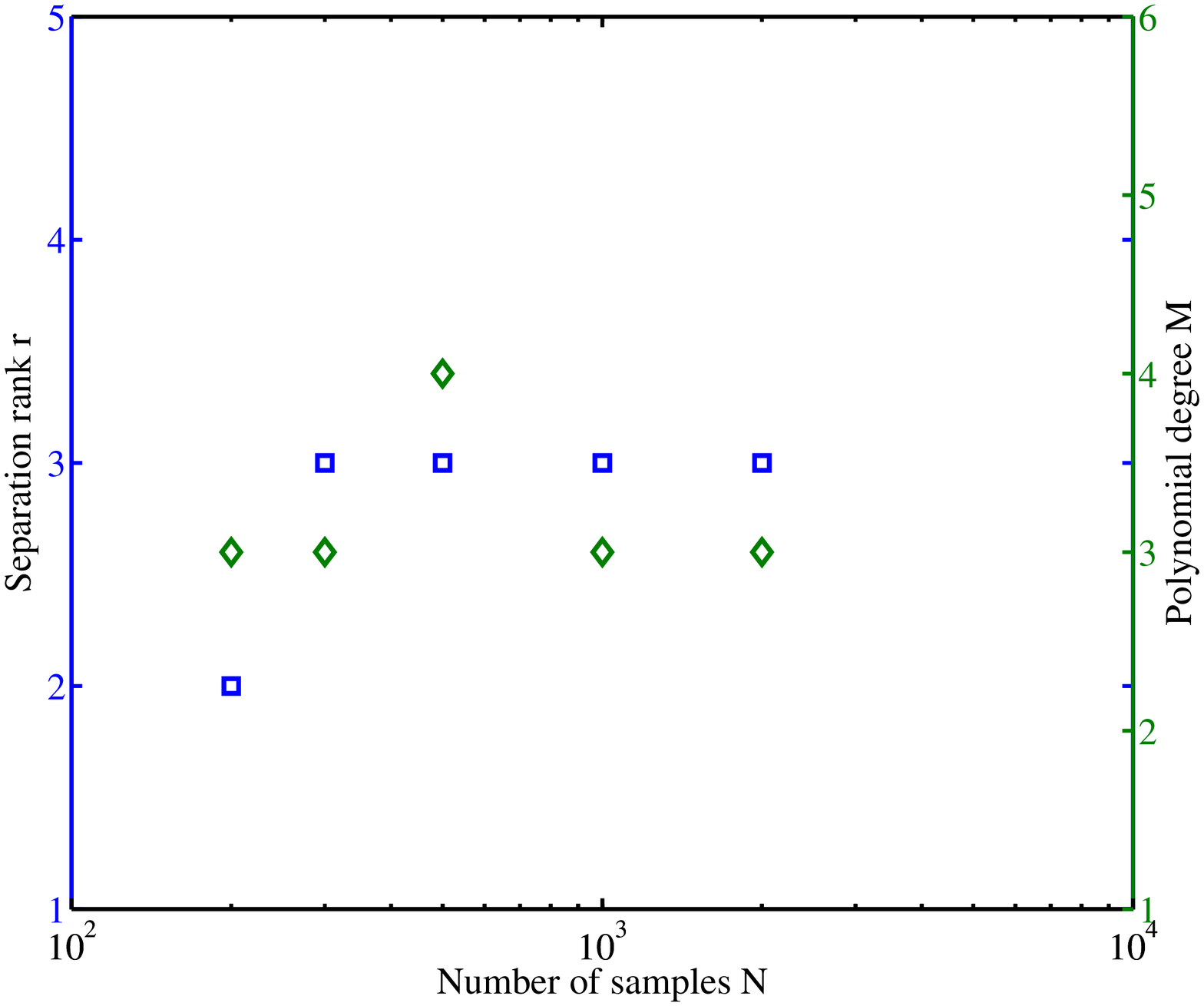}  
      &
      \includegraphics[width=2.8in]{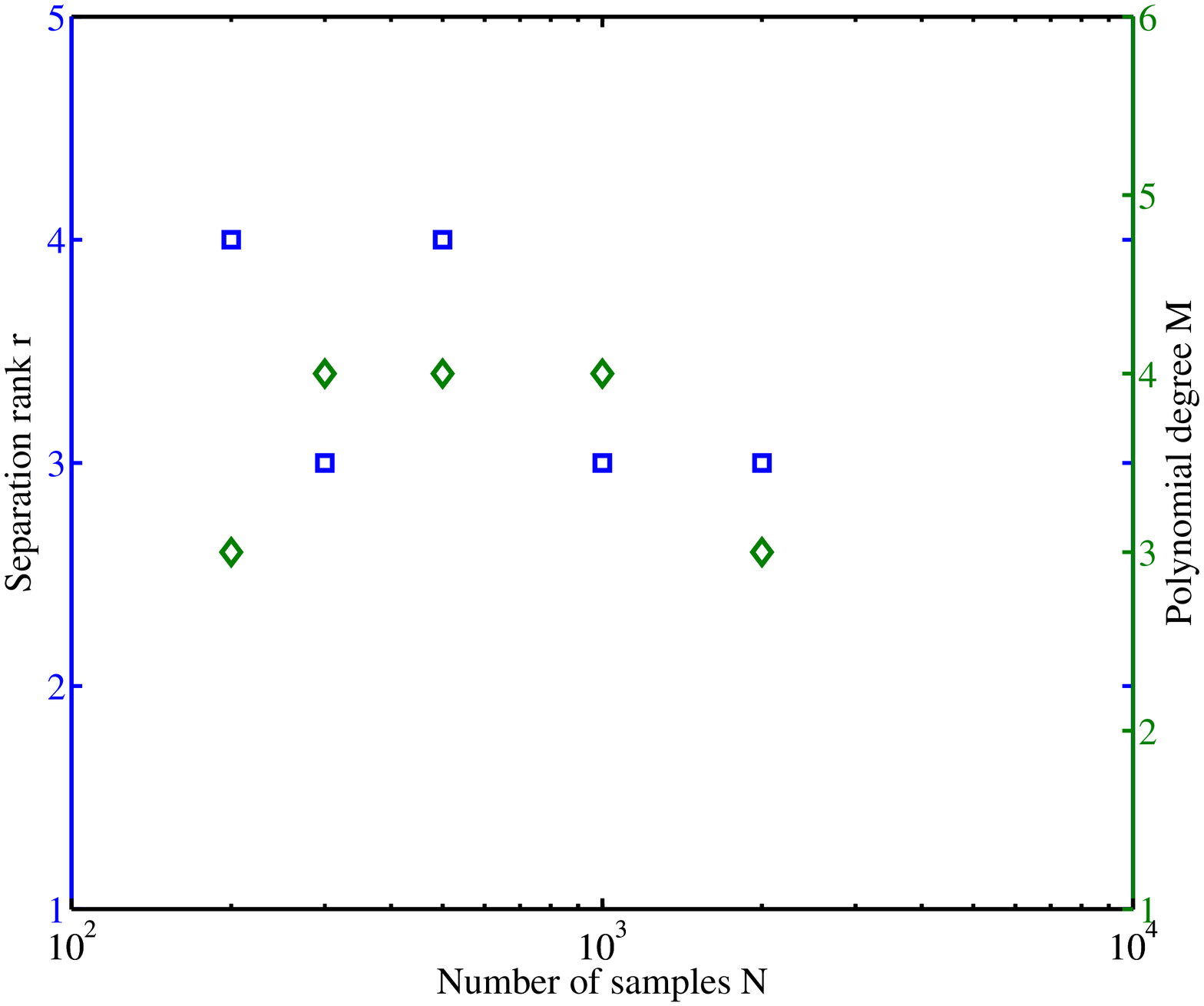} 
      \\
      (a) & (b)
      \\
      \includegraphics[width=2.8in]{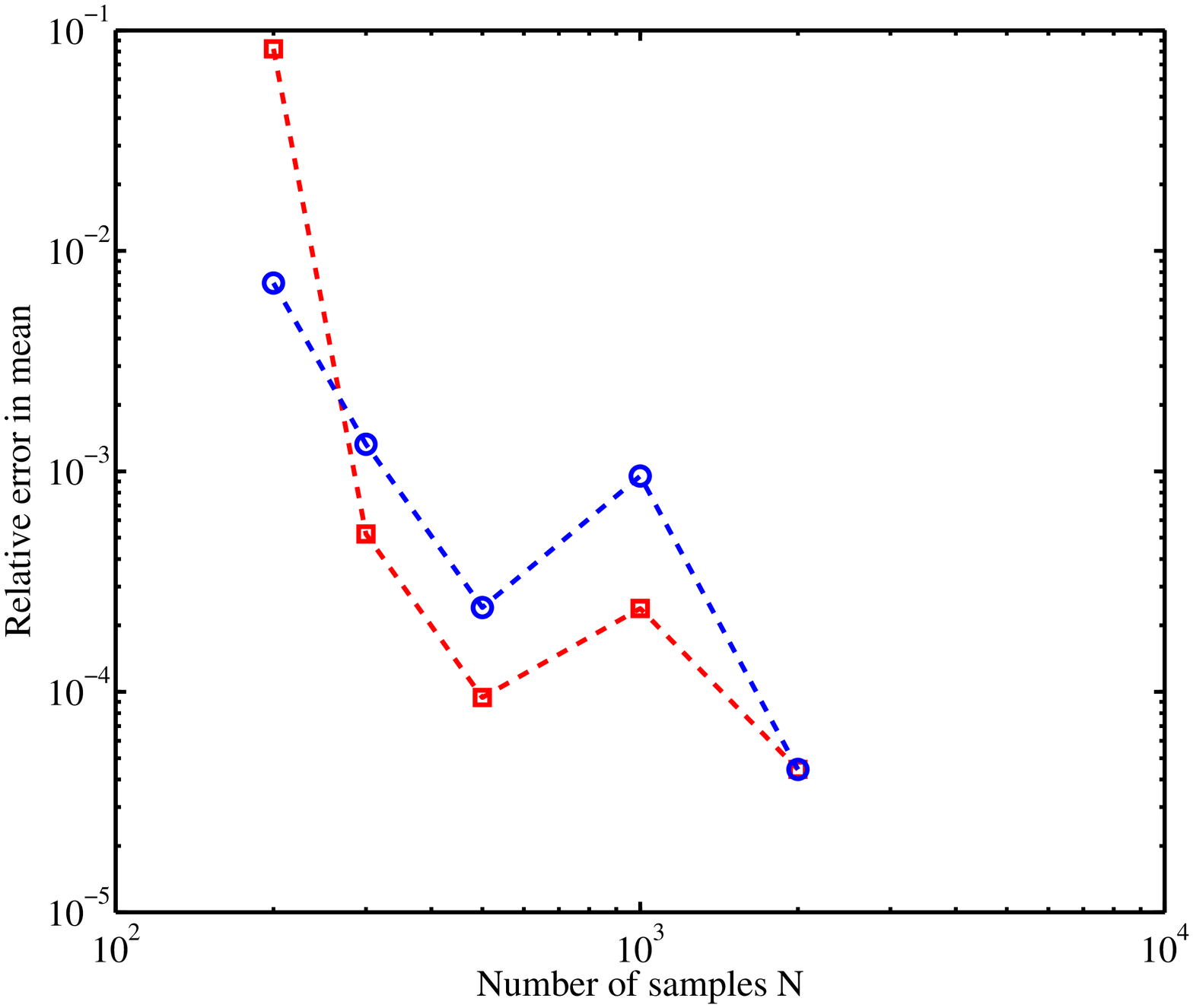}  
      &
      \includegraphics[width=2.8in]{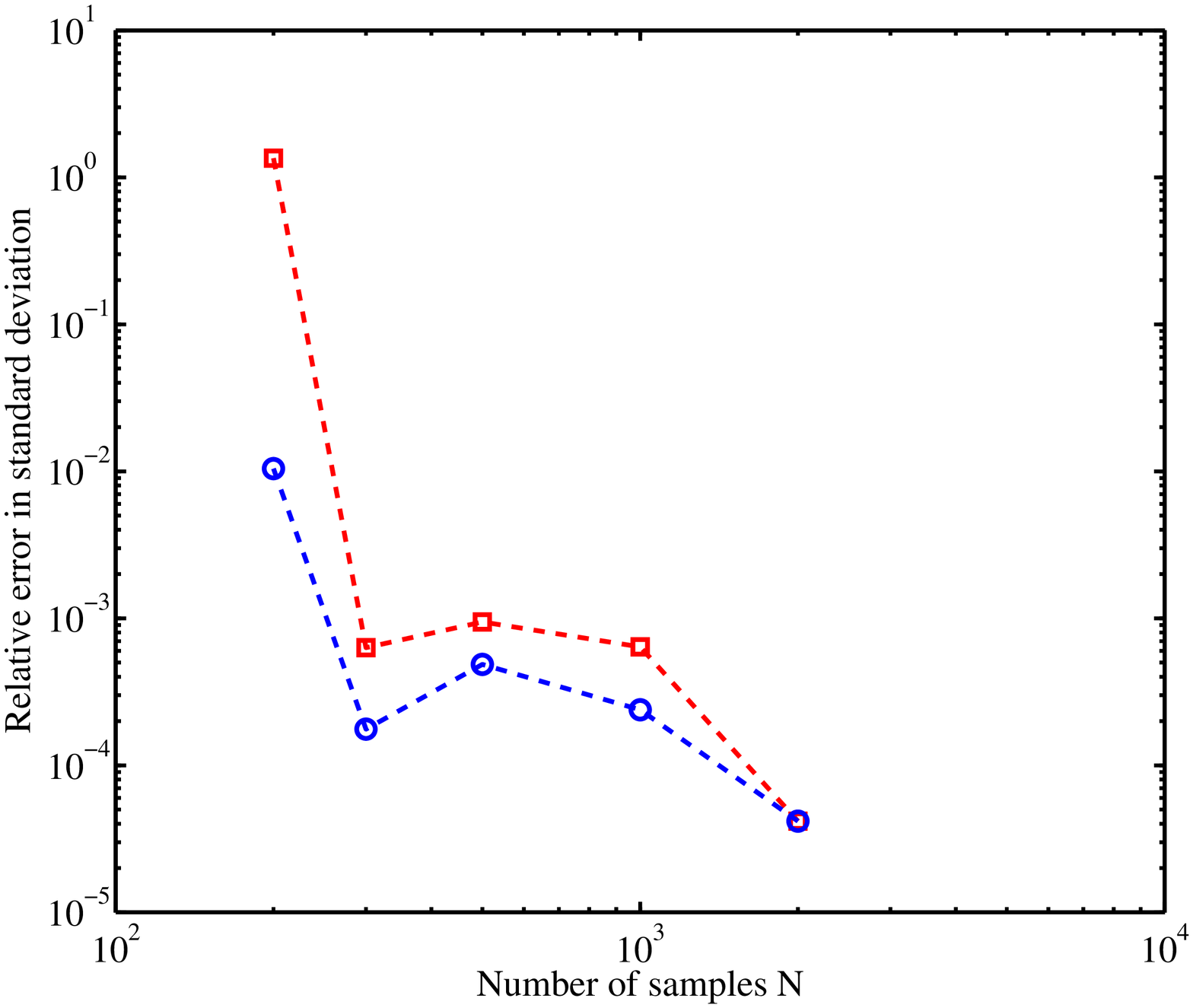}           
      \\
      (c) & (d)   
      
     \end{tabular}
      \caption{Evaluating the performance of the error indicator of Section \ref{sec:Perturbation_based_error} in estimating the separation rank $r$ and polynomial degree $M$. (a) $(r,M)$ estimated based on the error indicator of Section \ref{sec:Perturbation_based_error}; (b) $(r,M)$ that minimize the standard deviation error (computed based on the reference solution); (c) and (d) Mean and standard deviation error, respectively, when $r$ and $M$ are estimated using the error indicator and based on the minimum standard deviation error. (Separation rank $r$ ({\scriptsize $\square$}); Polynomial degree $M$ ($\diamond$); $(r,M)$ estimated based on the error indicator ({\scriptsize$\;\;\;\; \square$} $\hspace{-.73cm} ---$); $r$ and $M$ that minimize the standard deviation error ($\;\;\;\; \circ $ $\hspace{-.69cm} ---$)).}            
\label{fig:Manufactired_Optimum_rM}       
       \end{figure}

Figures \ref{fig:Manufactired_Optimum_rM}(a) and \ref{fig:Manufactired_Optimum_rM}(b) depict the selection of $(r,M)$ based on the error indicator of Section \ref{sec:Perturbation_based_error} and by minimizing the standard deviation error, respectively. In most cases the error indicator selects the reasonable values $r=3$ and $M=3$. As is observed from Figs. \ref{fig:Manufactired_Optimum_rM}(c) and \ref{fig:Manufactired_Optimum_rM}(d), except for the small sample size of $N=200$, the solution computed based on the error indicator leads to accuracies that are close to those generated based on minimizing the standard deviation error. 

%

\subsection{Linear elliptic stochastic equation}
\label{sec:Elliptic}

We consider the elliptic stochastic equation
\begin{eqnarray}\label{eq:1D elliptic}
-\frac{d}{dx} \left( a(x,\bm y) \frac{d u(x,\bm y)}{dx} \right)  &=&  1, \ \ \ \ x\in \mathcal{D}=\left(0,1\right),\\
u(0,\bm y) &=& u(1,\bm y)=0,\nonumber
\end{eqnarray}
where $a(x,\bm y)$ is the stochastic diffusion coefficient given by
\begin{equation}\label{eq:diffusion coefficient}
a(x,\bm y)=\bar{a}(x)+\sigma_a \sum_{k=1}^{d} \sqrt{\gamma_k} \phi_k(x)y_k. \nonumber
\end{equation}

Here, $\bar{a}(x)$ is the mean of $a(x,\bm y)$ and $\sigma_a$ is a coefficient to control the variability of $a(x,\bm y)$. The random variables $\{y_k\}_{k=1}^d$ are assumed to be independent  and uniformly distributed on $[-1,1]$. Additionally, $\{\gamma_i\}_{k=1}^{d}$ and $\{\phi_k(x)\}_{k=1}^{d}$ are, respectively, $d$ largest eigenvalues and the corresponding eigenfunctions of the covariance function
\begin{equation}\label{eq:Gaussian correlation}
C_{aa}\left(x_1,x_2 \right)=\exp \left[-\frac{\left(x_1-x_2\right)^2}{l_c^2} \right],\quad x_1,x_2\in \mathcal{D},\nonumber
\end{equation}
where $l_c$ is the correlation length of $a(x,\bm y)$ and is set to $l_c=1/14$. This will lead to $d=40$ dominant eigenvalues in the spectrum of $C_{aa}$. Furthermore, we consider $\overline{a}=0.1$ and choose $\sigma_a=0.021$ to guarantee the positivity of $a(x,\bm y)$ over $\mathcal{D}$. The realization of $u$ in (\ref{eq:1D elliptic}) are obtained by a Finite Element solver with quadratic elements. The mean and standard deviation of the solution at $x=0.5$ are the quantities of interest. The accuracy of the non-intrusive separated representation is compared against those of the Monte Carlo simulation and sparse grid stochastic collocation techniques. To obtain a reference solution, we compute the $3$rd order PC expansion of solution at $x=0.5$ using level $l=5$ stochastic collocation with the Clenshaw-Curtis rule (see, e.g., \cite{Xiu05a} for more information on the stochastic collocation approach and the Clenshaw-Curtis rule). 

\begin{figure} 
    \centering
    \begin{tabular}{cc}
            \hspace{-0.5cm}    
      \includegraphics[width=2.8in]{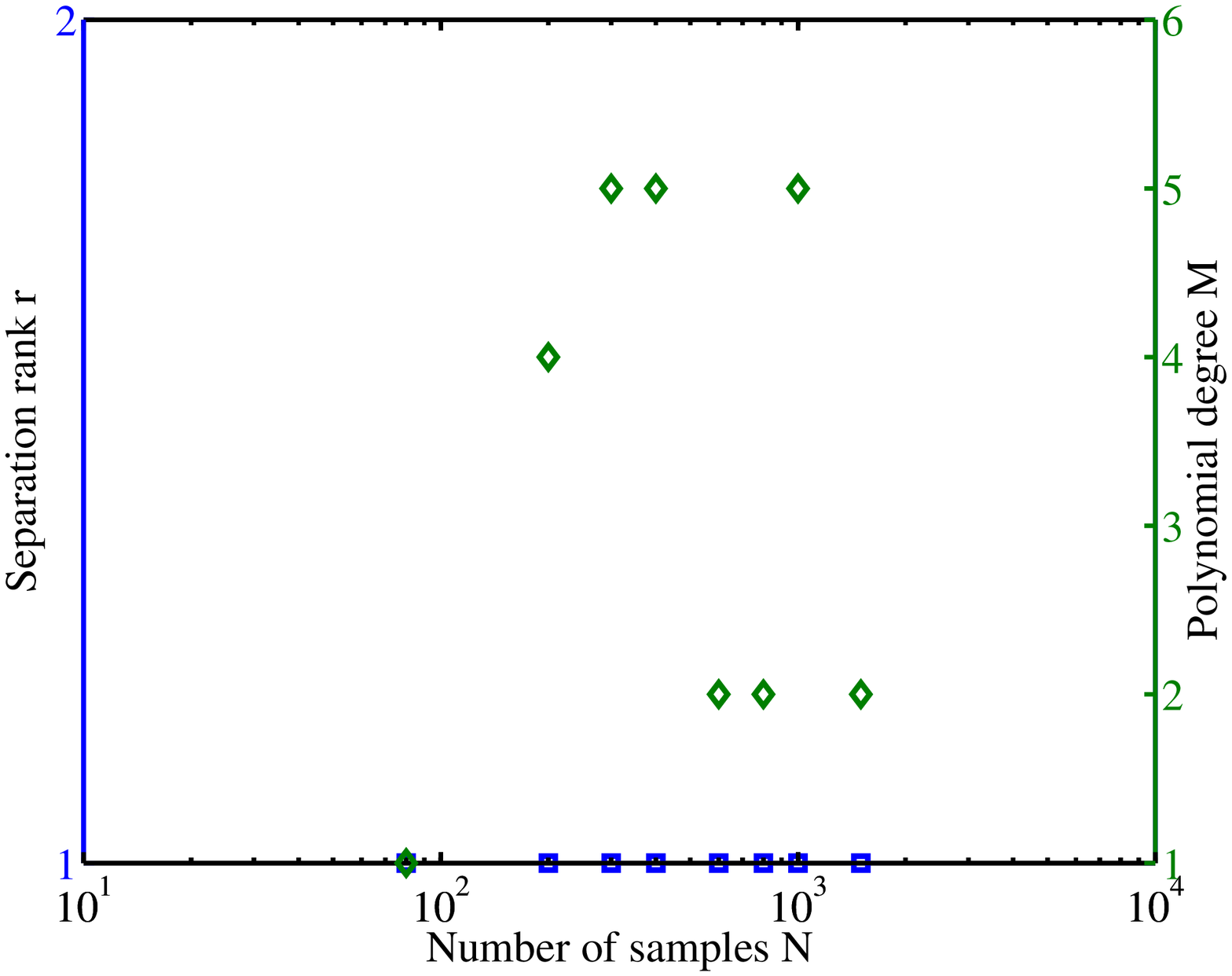}  
      &
      \includegraphics[width=2.8in]{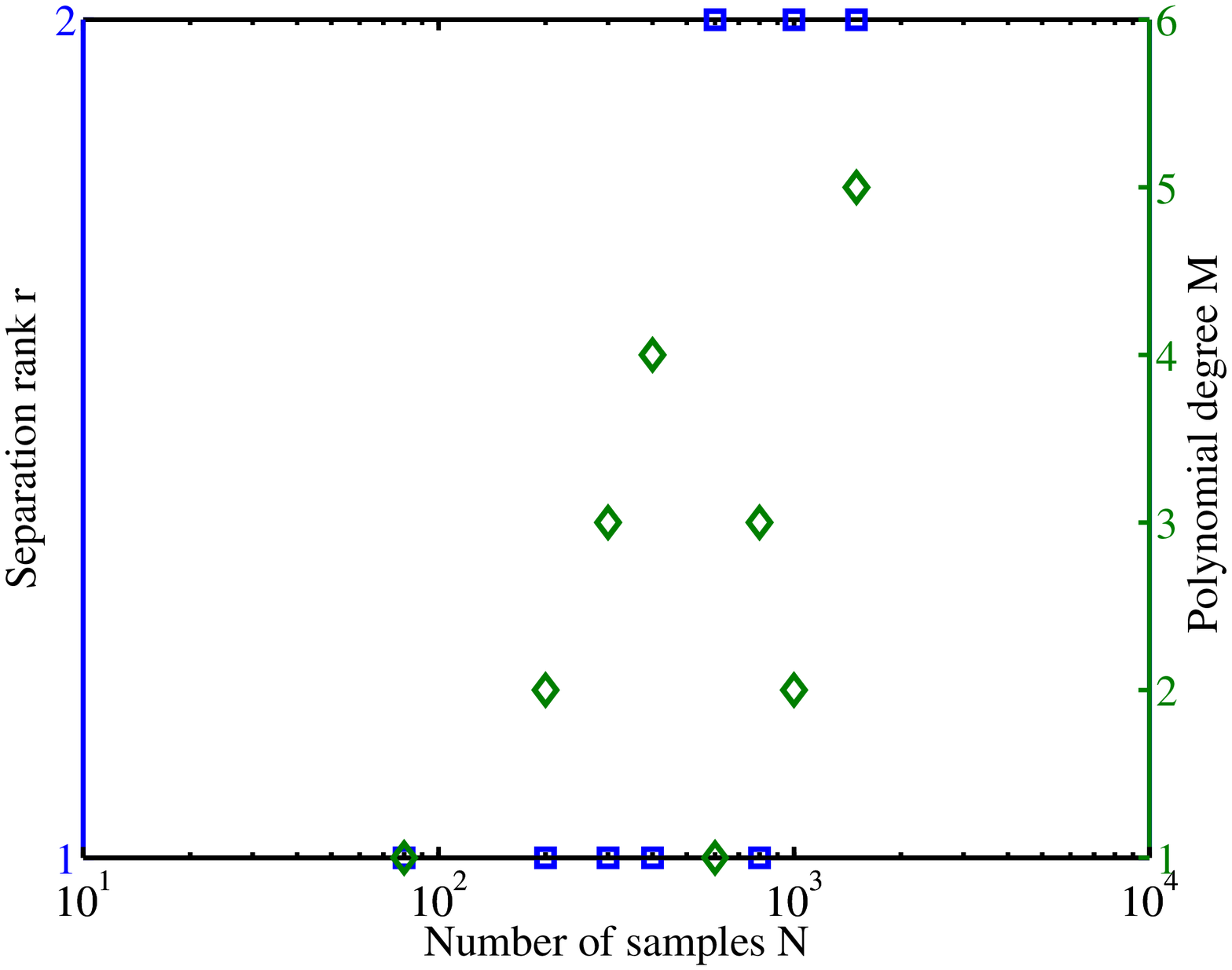} 
      \\
      (a) & (b)
      \\
      \includegraphics[width=2.8in]{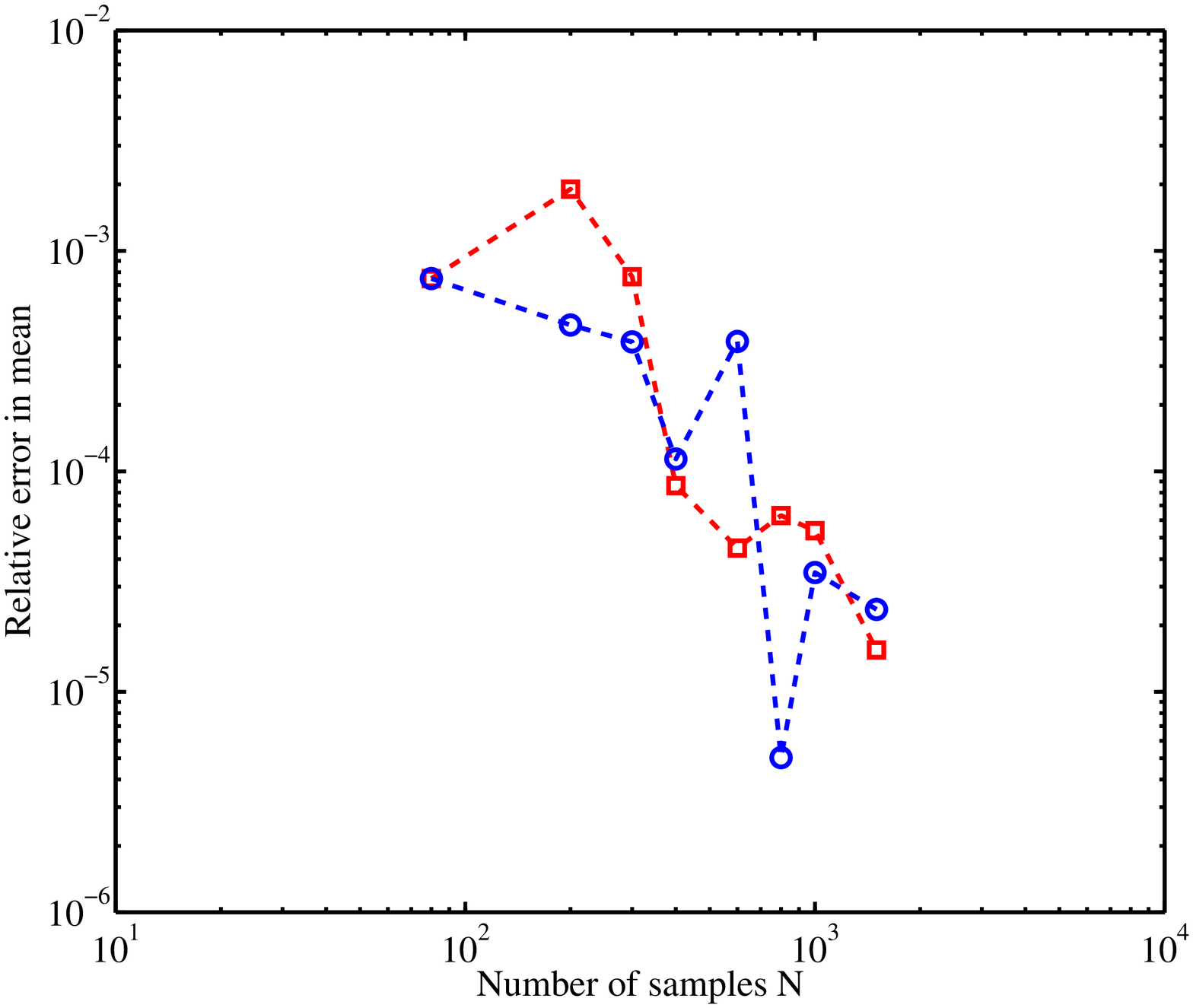}  
      &
      \includegraphics[width=2.8in]{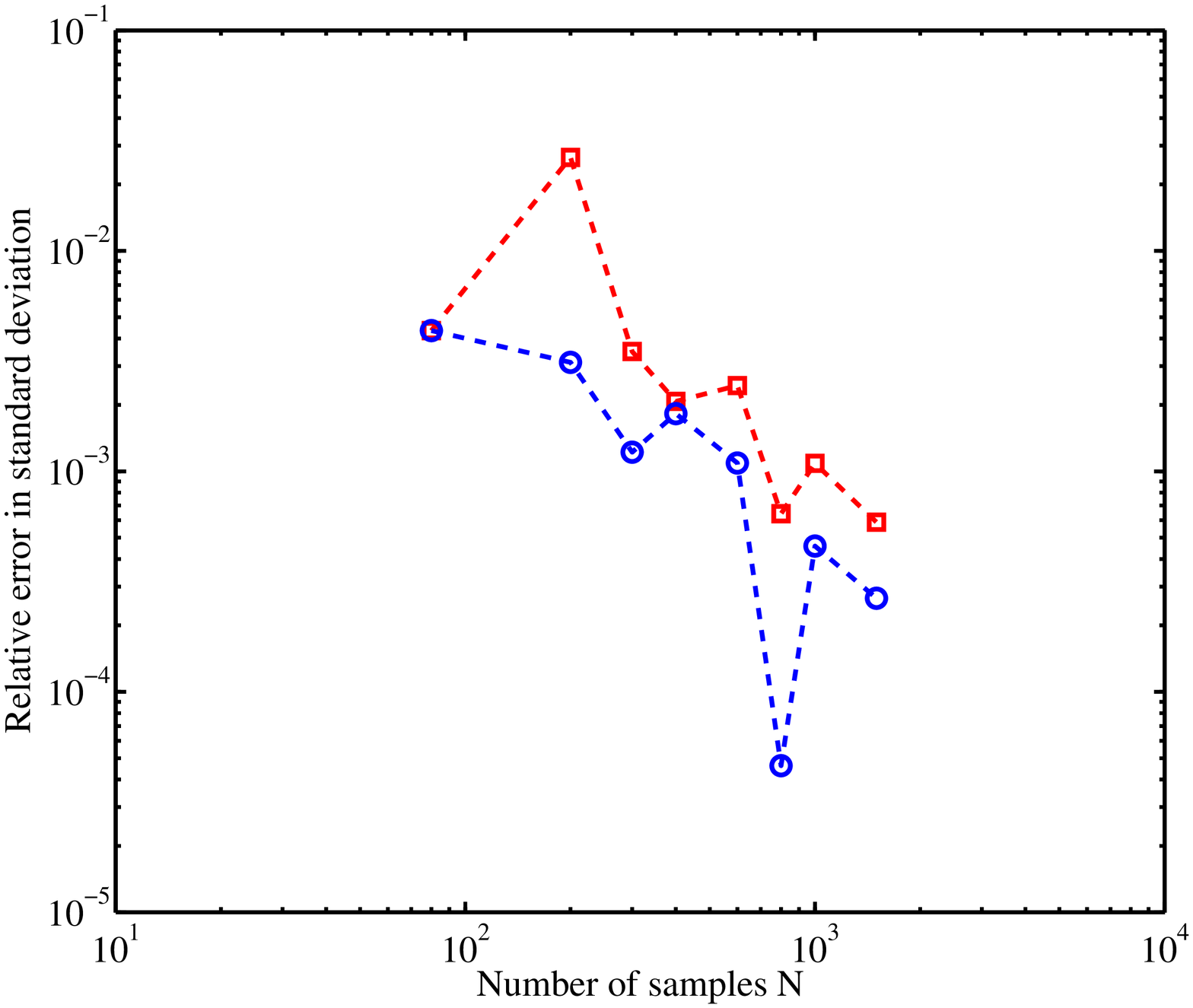}           
      \\
      (c) & (d)   
      
     \end{tabular}
      \caption{Evaluating the performance of the error indicator of Section \ref{sec:Perturbation_based_error} in estimating the separation rank $r$ and polynomial degree $M$. (a) $(r,M)$ estimated based on the error indicator of Section \ref{sec:Perturbation_based_error}; (b) $r$ and $M$ that minimize the standard deviation error (computed based on the reference solution); (c) and (d) Mean and standard deviation error, respectively, when $r$ and $M$ are estimated using the error indicator and based on the minimum standard deviation error. (Separation rank $r$ ({\scriptsize $\square$}); Polynomial degree $M$ ($\diamond$); $(r,M)$ estimated based on the error indicator ({\scriptsize$\;\;\;\; \square$} $\hspace{-.73cm} ---$); $(r,M)$ that minimize the standard deviation error ($\;\;\;\; \circ $ $\hspace{-.69cm} ---$)).}            
\label{fig:Elliptic Optimum rM yy}       
       \end{figure}

Given the uniform distribution of $y_k$, we expand the univariate factors $\{u_k^l(y_k)\}$ in a Legendre polynomial basis. Figure \ref{fig:Elliptic Optimum rM yy}(a) displays the estimates of the separation rank $r$ and polynomial degree $M$ obtained for sample sizes $N \in\{ 80,200,300,400,600,800,1000,1500 \}$ using the error indicator $EI$ discussed in Section \ref{sec:Perturbation_based_error}. As can be observed from this figure,  $r=1$ is selected for different values of $N$, thus confirming that $u(0.5,\bm y)$ admits a low-rank separated representation. However, the estimates of the polynomial degree $M$ depend on $N$. In Fig. \ref{fig:Elliptic Optimum rM yy}(b), for each $N$, we obtain the pair $(r,M)$ such that the resulting error in the standard deviation of the solution is minimum. Notice that such a selection of $(r,M)$ requires the availability of the reference solution. Figures \ref{fig:Elliptic Optimum rM yy}(c) and \ref{fig:Elliptic Optimum rM yy}(d) illustrate the convergence of the solution mean and standard deviation corresponding to $(r,M)$ selected according to the above two strategies. While the estimates of $(r,M)$ based on the error indicator are not precisely the same as those minimizing the standard deviation error, the resulting accuracies in Figs. \ref{fig:Elliptic Optimum rM yy}(c) and \ref{fig:Elliptic Optimum rM yy}(d) are comparable. Moreover, the $EI$-based selection of $(r,M)$ leads to stable separated representations of the solution.

\begin{figure}
    \centering
    \begin{tabular}{c}
            \hspace{-0.3cm}    
      \includegraphics[width=2.9in]{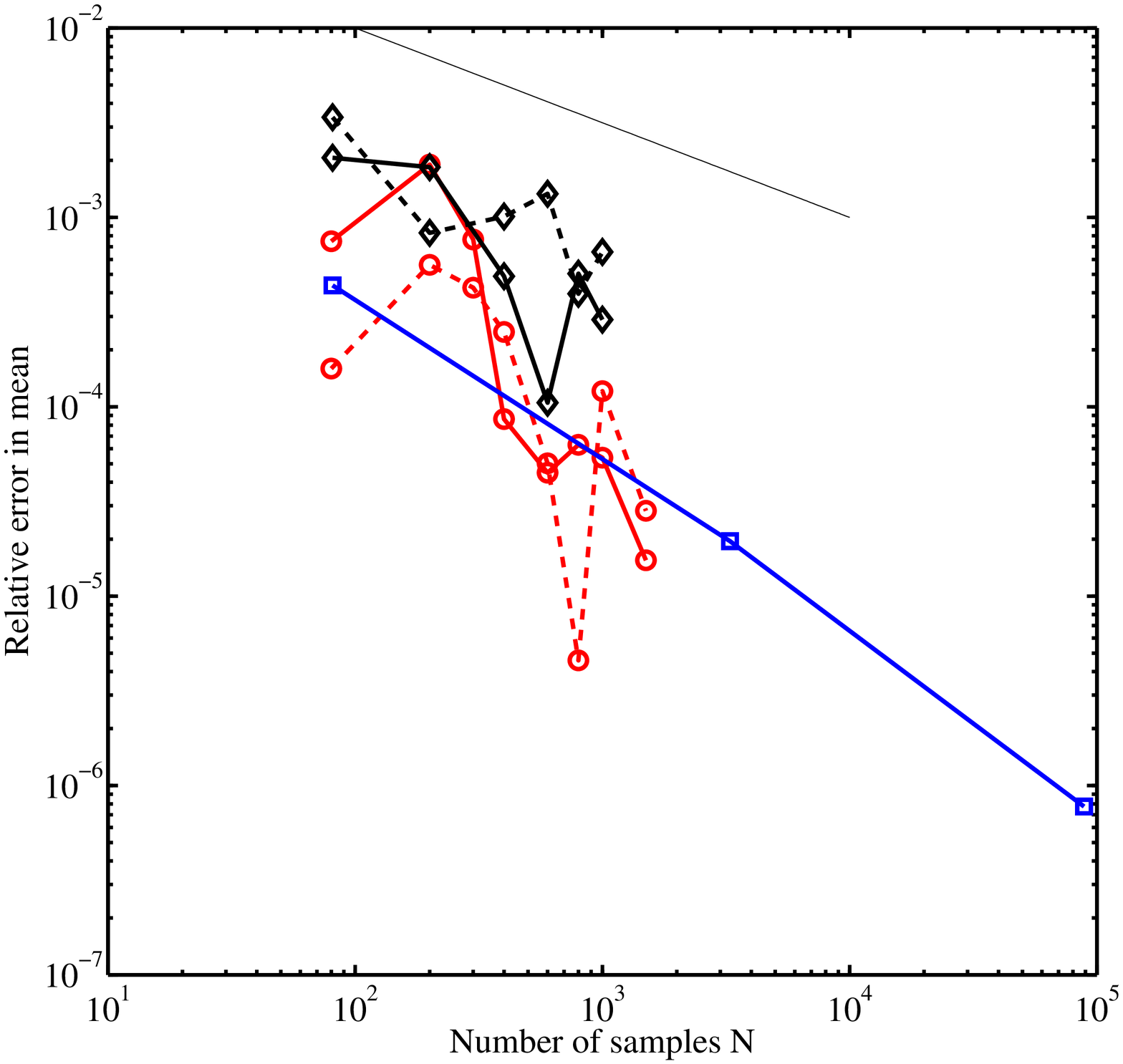}
     
            \hspace{-.5cm}
      \includegraphics[width=2.9in]{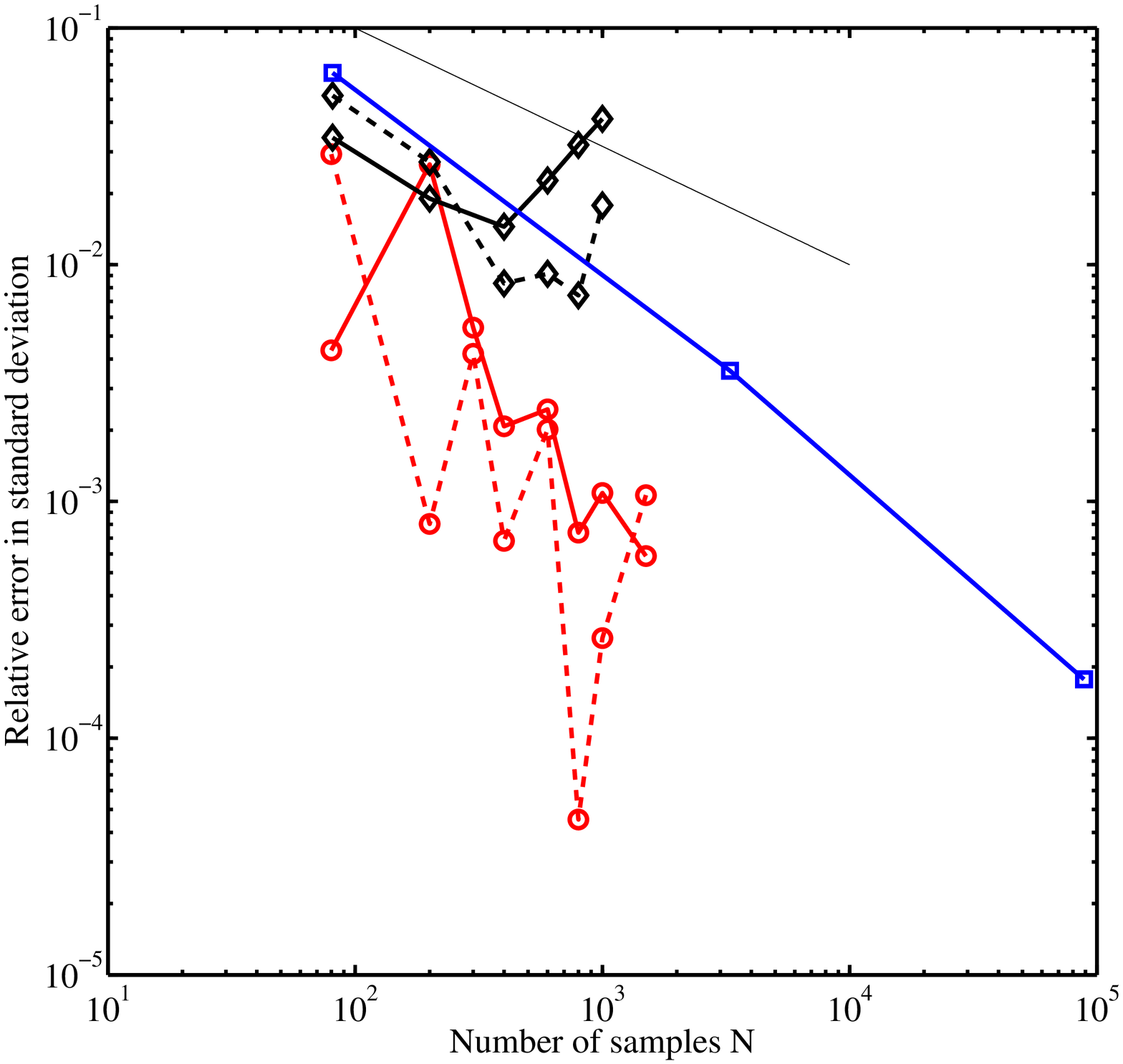}
      \\
       \put(85,0){\footnotesize(a)}
        \put(300,0){\footnotesize(b)}
          \\
     \end{tabular}
      \caption{Comparison of relative mean and standard deviation errors of the solution at $x=0.5$ for the separated representation, Monte Carlo simulation, and sparse grid stochastic collocation with Clenshaw-Curtis abscissas (First set: solid line; Second set: dashed line). (a) Relative error in mean and (b) Relative error in standard deviation. (Monte Carlo simulation ($\diamond$); Stochastic collocation ({\scriptsize $\square$}); Separated representation($\circ$); $1/\sqrt{N}$ decay rate (Thin continuous line \line(1,0){15})).}            
 \label{fig:elliptic_error}     
       \end{figure}

Figures \ref{fig:elliptic_error}(a) and \ref{fig:elliptic_error}(b) compare the convergence of the  mean and standard deviation of $u(0.5,\bm y)$ obtained by the separated representation, the standard Monte Carlo simulation, as well as the (isotropic) sparse grid stochastic collocation with the Clenshaw-Curtis abscissas. As the estimates obtained by the separated representation and the Monte Carlo simulation are sample dependent, two sets of independent realizations are considered. 
Figure \ref{fig:elliptic_error}(b), in particular, demonstrates a faster convergence and more accurate estimates of the solution standard deviation achieved by the separated representation approach. Additionally, as the construction of the non-intrusive separated representation is based on random sampling of the solution, the approximation refinement may be achieved by incorporating as many additional samples as can be afforded. In the sparse grid stochastic collocation with nested abscissas, however, the number of additional solution samples is dictated by the underlying quadrature rule. 

\subsection{The hydrogen oxidation problem}
\label{sec:Hydrogen Oxidation Problem}

As a third test case, we study the problem of hydrogen oxidation under the SuperCritical Water (SCW) condition. SCW condition provides a means for fast and complete oxidation of organics without toxic formation. The chemical mechanism associated with the hydrogen oxidation may involve numerous uncertain parameters such as coefficients describing elementary reactions and thermodynamic properties of species, and has been the subject of several numerical studies in the area of uncertainty quantification \cite{Phenix98,Reagan03,Reagan04,Reagan05,LeMaitre07b,Mathelin10,Alexanderian11}.

In the present work, we focus on the reduced mechanism of \cite{Phenix98} consisting of $8$ reversible elementary reactions and $7$ species: $OH$, $H$, $H_2 O$, $H_2 $, $O_2 $, $HO_2 $, and $H_2 O_2$. We assume that forward reaction rate constants $k_{f,j}$ and species enthalpies of formation $\Delta H^{0}_{f,i}$ (associated with $OH$, $H$, $H_2 O$, $HO_2 $, and $H_2 O_2$) are independent uncertain parameters, thus giving rise to an uncertainty space of dimensionality $d=13$. In particular, following \cite{Phenix98,Reagan03,Reagan04}, we model $k_{f,j}$ as independent lognormal random variables with prescribed medians $\tilde k_{f,j}$ and such that $Prob\ [\tilde k_{f,j}/\zeta_j\le k_{f,j} \le \tilde k_{f,j} \zeta_j ] = 1 - \theta$. The so-called {\it uncertainty factors} $\zeta_j$ are given in Table \ref{table:mechanism}, and we here set $\theta = 0.004$.  The medians $\tilde k_{f,j}$ are modeled by the Arrhenius equation $\tilde k_{f,j}=A_jT^{n_j} e^{-E_{a,j}/RT}$, where $R$ is the universal gas constant, $T$ is the temperature, and parameters $A_j$, $E_{a,j}$, and $n_j$ are derived from experimental data, see Table \ref{table:mechanism}. The realizations of $k_{f,j}$ with the above descriptions may be obtained from 
\begin{equation}
\label{eqn:forward_enth}
k_{f,j} = \tilde k_{f,j}\ \exp\left(\frac{\log \zeta_j}{\Phi^{-1}(1-\frac{\theta}{2})}\ y_{j}\right)\ , \quad j=1,\dots,8,\nonumber
\end{equation}
in which $\{y_{j}\}_{j=1}^8$ are independent standard Gaussian random variables with cumulative distribution function (CDF)  $\Phi$. Moreover, we characterize the uncertainty in the species enthalpies of formation $\Delta H^{0}_{f,i}$ as independent Gaussian random variables with experimental mean values and standard deviations reported in Table \ref{table:musigma}. Other model parameters including entropies, internal energy, and initial conditions are considered to be deterministic. \\[-.3cm]


%

\begin{table}[h]
\caption{Reduced mechanism of hydrogen oxidation \cite{Phenix98}. $A_j$: Pre-exponential coefficient, $n_j$: Temperature exponent, $E_{a,j}/R$: Activation energy over universal gas constant, $\zeta_j$: Uncertainty factor.
} 
\centering
\begin{tabular}{ c c c c c c }   
\hline 
$j$ & Elementary reaction  & $A_j$ & $n_j$ & $E_{a,j}/R$ & $\zeta_j$ \\ 
\hline\hline 
1 & $OH$ + $H$ $\longleftrightarrow$ $H_2O$ & $1.620E^{+14}$ & 0 & 75 & 3.16 \\ 
2 & $H_2$ + $OH$ $\longleftrightarrow$ $H_2O$ + $H$ & $1.024E^{+8}$ & 1.6 & 1660 & 1.26 \\ 
3 & $H$ + $O_2$  $\longleftrightarrow$ $HO_2$ & $1.481E^{+12}$ & 0.6 & 0 & 1.58 \\ 
4 & $HO_2$ + $HO_2$  $\longleftrightarrow$ $H_2O_2$ + $O_2$ & $1.620E^{+12}$ & 0 & 775 & 1.41 \\ 
5 & $H_2O_2$ + $OH$ $\longleftrightarrow$ $H_2O$  + $HO_2$ & $1.789E^{+12}$ & 0 & 670 & 1.58 \\ 
6 & $H_2O_2$ + $H$  $\longleftrightarrow$ $HO_2$ + $H_2$ & $1.686E^{+12}$ & 0 & 1890 & 2.00 \\ 
7 & $H_2O_2$  $\longleftrightarrow$ $OH$ + $OH$ & $3.000E^{+14}$ & 0 & 24400 & 3.16 \\ 
8 & $OH$ + $H_2O$  $\longleftrightarrow$ $H_2O$ + $O_2$ & $2.891E^{+13}$ & 0 & -250 & 3.16 \\ 
\hline 
\end{tabular} 
\label{table:mechanism} 
\end{table}


\begin{table}[h] 
\caption{Means and standard deviations of the species enthalpies of formation $\Delta H^{0}_{f,i}$ \cite{Phenix98}.}
\centering
\begin{tabular}{ c c c c c c }
Species & $H$ & $OH$ & $H_2O$ & $H_2O_2$ & $HO_2$ \\ 
\hline\hline
index ($i$) & 1 & 2  &  3 & 4 & 5\\
$\mu_i$ & 52.10 & 9.3 & -57.80 & -32.53 & 3.00 \\ 
$2\sigma_i$ & 0.01 & 0.2 & 0.01 & 0.07 & 0.5 \\ 
\hline 
\end{tabular}
\\ 
\label{table:musigma}
\end{table}

%
\noindent{\it Perfectly stirred reactor (PSR).} We process the oxidation in a perfectly stirred reactor (PSR), where the combustion is assumed to be spatially homogeneous as a consequence of high diffusion rates and forced turbulent mixing. The PSR (isochoric here) is modeled by the conservation of mass, energy, and species, and we here use the CHEMKIN code \cite{Kee93} to solve the governing equations over a $10\ sec$ time interval. The deterministic initial thermodynamic conditions are $T=823\ kelvin$ and $P=246\ bar$, with initial species concentration $C_{H_2 }=0.481\times10^{-3}\ \frac{mol}{cm^3}$, $C_{O_2 }=0.243\times 10^{-3}\ \frac{mol}{cm^3}$ , and $C_{H_2O }=0.99\times 10^{-3}\ \frac{mol}{cm^3}$. For further information about the details of this problem, we refer the interested reader to \cite{Phenix98}.

%

%
\subsubsection{Uncertainty quantification of $OH$ concentration}
\label{sec:Results of Hydrogen Oxidation}

We apply the proposed non-intrusive separated approximation to compute the $OH$ concentration as well as its mean and standard deviation. In order to assess the quality of our approximations, we generate an order $p=4$ Hermite PC reference solution whose coefficients are obtained using least-squares regression with $N=100,000$ realizations of $OH$ concentration. In particular, we investigate the effect of two choices of $\bm L$ as well as the performance of the proposed error indicator for the selection of separation rank $r$ and the Hermite polynomial degrees $M$. Additionally, we compare the accuracy of the solutions obtained using separated representation and the regression-based PC for small values of $N$. 

\begin{figure} 
    \centering
    \begin{tabular}{c}
            \hspace{-0.5cm}    
\begin{tiny}

\end{tiny}    
  \includegraphics[width=3.5in]{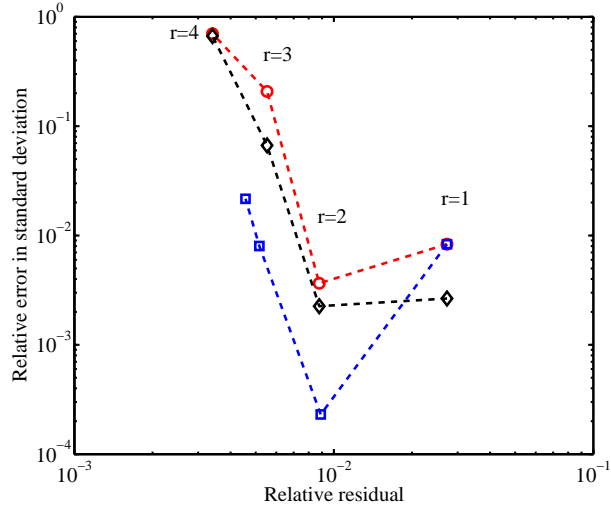}
     
     \end{tabular}
      \caption{ Regularized and non-regularized estimation of solution standard deviation as a function of the relative residual $\Vert u - u_r\Vert_D/\Vert u\Vert_D$ ($N=200$, $t=2.2\ sec$, and $M=2$). 
For separation ranks $r>2$ where over-fitting occurs, the Tikhonov matrix of Section \ref{sec:tikh_matrix} is considerably more effective than regularizing with  $\bm L = \bm{I}_{M+1}\otimes diag(s_1,\dots,s_r)$. (Regularized with $\bm L$ defined in Section \ref{sec:tikh_matrix} ({\scriptsize $\square$}); Regularized with $\bm L = \bm{I}_{M+1}\otimes diag(s_1,\dots,s_r)$ ($\diamond$); Non-regularized case($\circ$)).}            
\label{fig:regularization effects}      
       \end{figure}

Figure \ref{fig:regularization effects} compares the relative error in estimating the standard deviation of $OH$ concentration at $t=2.2\ sec$ with and without the Tikhonov regularization. For the regularized case, the Tikhonov matrix $\bm L$ of Section \ref{sec:tikh_matrix} as well as $\bm L = \bm{I}_{M+1}\otimes diag(s_1,\dots,s_r)$ were used. As can be observed from Fig. \ref{fig:regularization effects}, for large separation ranks $r$, the method over-fits the data. More importantly, the regularized approximation with the Tikhonov matrix $\bm L$ of Section \ref{sec:tikh_matrix} considerably reduces the over-fitting, as a result of which higher accuracies are achieved compared to a non-regularized case or when $\bm L = \bm{I}_{M+1}\otimes diag(s_1,\dots,s_r)$ is applied. We refer to Section \ref{sec:tikh_matrix} for an explanation of this observation. 

\begin{figure}
    \centering
    \begin{tabular}{c}
            \hspace{-0.5cm}    
      \includegraphics[width=2.8in]{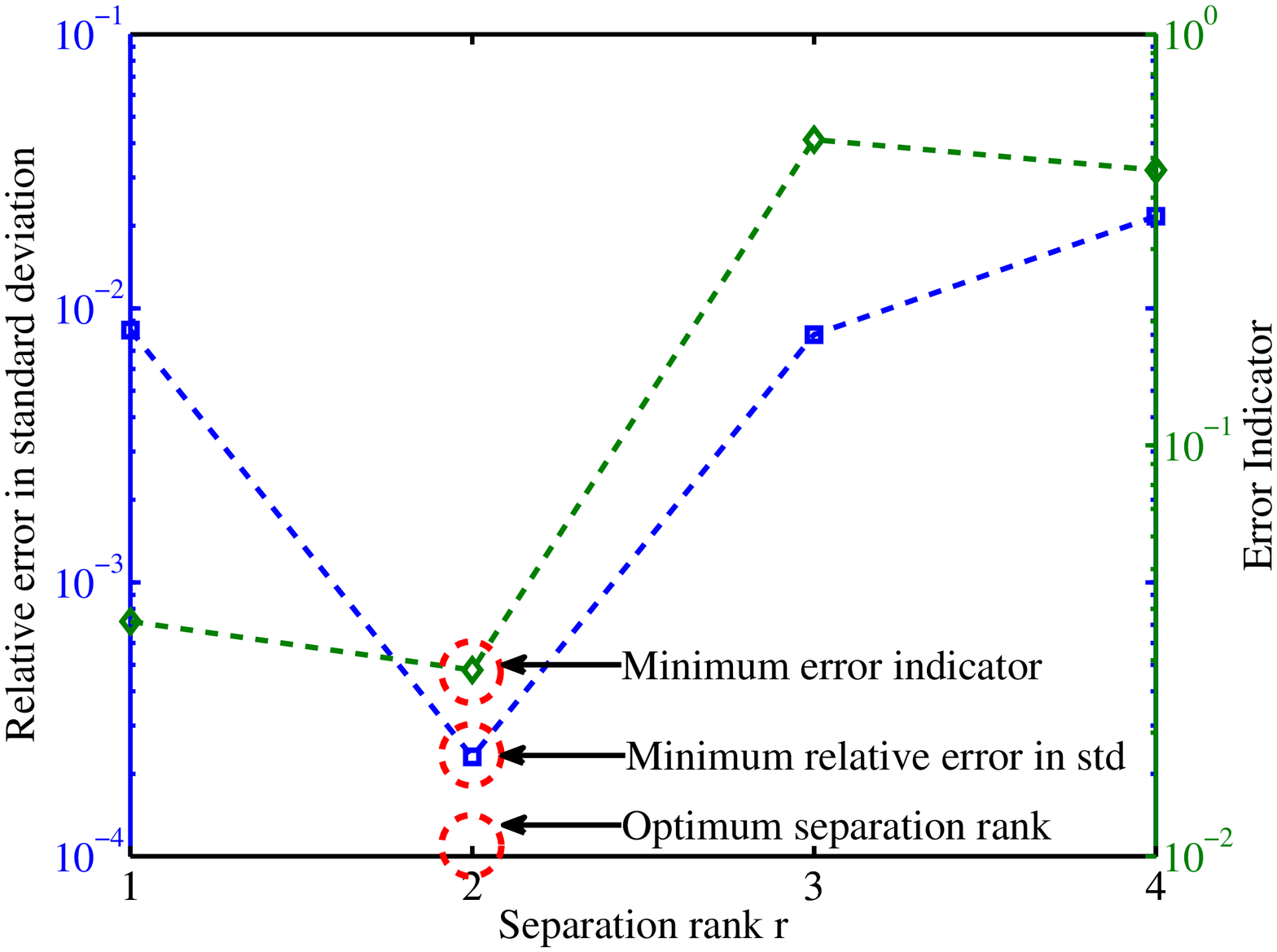}	     
      \hspace{.5cm}
      \includegraphics[width=2.8in]{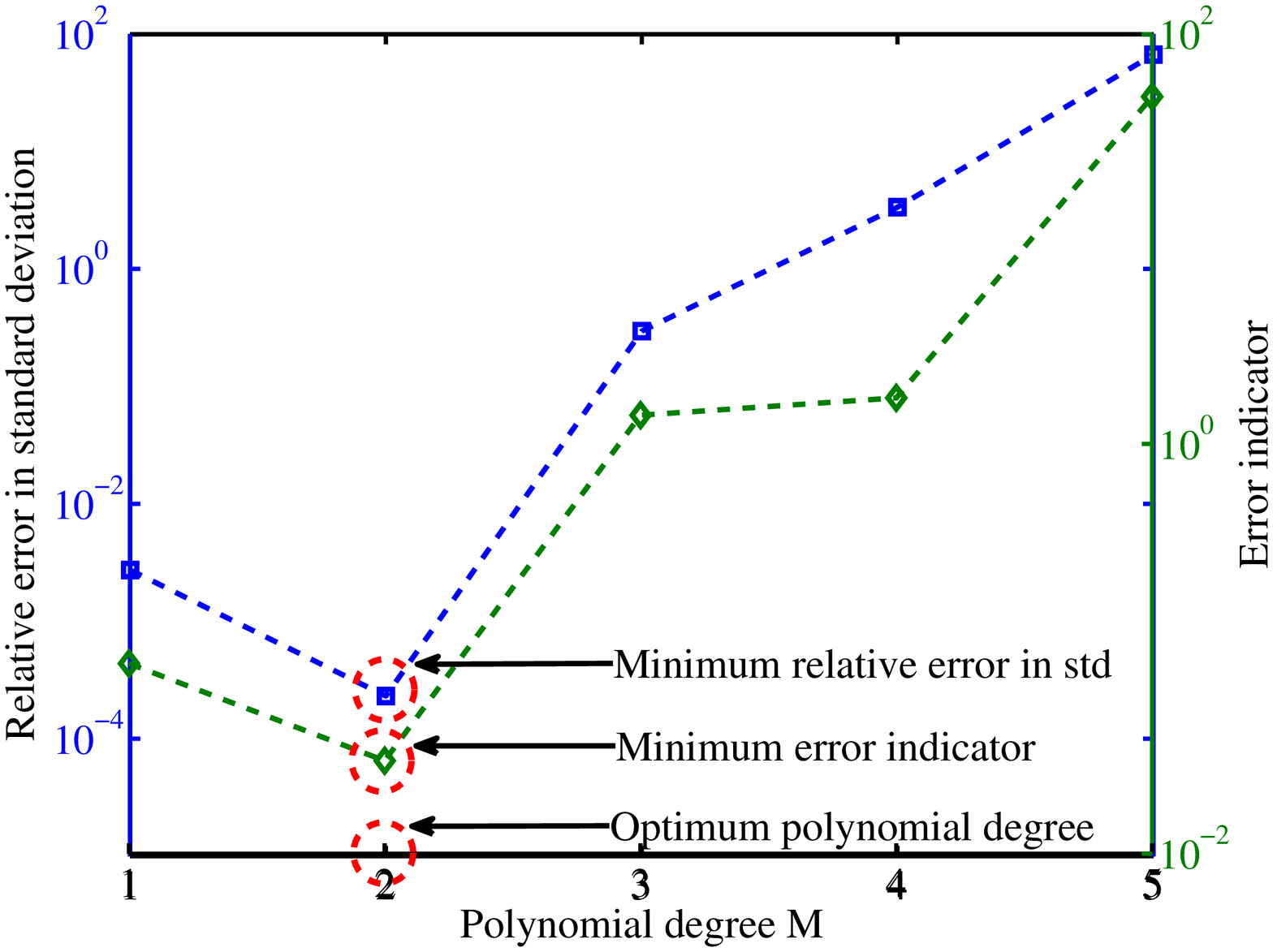}
       \\
       \put(85,0){\footnotesize(a)}
        \put(300,0){\footnotesize(b)}
          \\
     \end{tabular}
      \caption{Selection of the separation rank $r$ and spectral polynomial degree $M$ ($N=200$, $t=2.2\ sec$) using the error indicator of Section \ref{sec:Perturbation_based_error}. (a) Standard deviation error and the error indicator vs. $r$ for $M= 2$ and (b) Standard deviation error and the error indicator vs. $M$ for $ r= 2$. Here optimal $r$ and $M$ refer to those that correspond to smallest standard deviation error. (Standard deviation error ({\scriptsize $\square$}); Error indicator ($\diamond$)).}            
 \label{fig:Error Indicator}      
       \end{figure}

The performance of the error indicator of Section \ref{sec:Perturbation_based_error} in estimating the optimal $r$ and $M$ is illustrated in Fig. \ref{fig:Error Indicator} for the case of $N=200$ and $t=2.2\ sec$. Here, optimal $r$ and $M$ refer to those that correspond to smallest standard deviation error. In Fig. \ref{fig:Error Indicator}(a), we show the error indicator values as a function of $r$ while fixing $M=2$. In this case, the error indicator achieves its minimum at $r=2$, which happens to be the separation rank corresponding to the minimum standard deviation error. We stress that the standard deviation errors were generated based on the reference values, while the error indicator does not require the knowledge of the reference solution. In Fig. \ref{fig:Error Indicator}(b), we show a similar analysis except that we fix $r=2$ and let $M$ vary. 

\begin{figure} 
    \centering
    \begin{tabular}{c}
            \hspace{-0.5cm}    
      \includegraphics[width=3.5in]{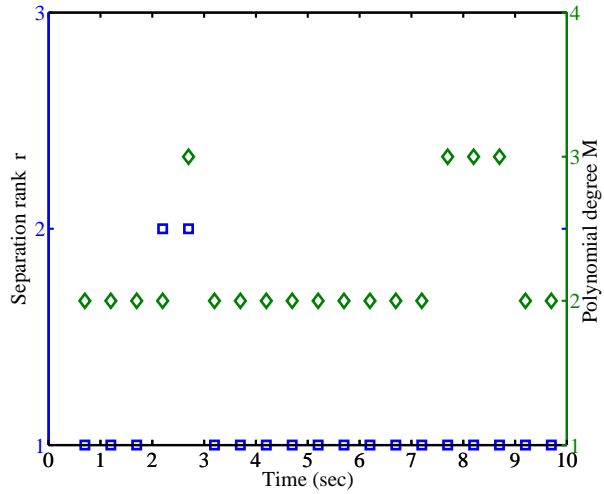}     
          \\
     \end{tabular}
      \caption{Values of separation rank $r$ and spectral polynomial degree $M$ ($N=200$) obtained from the error indicator of Section \ref{sec:Perturbation_based_error}. ($r$ ({\scriptsize $\square$}); $M$ ($\diamond$)).}            
\label{fig:Optimum r M}     
       \end{figure}

Figure \ref{fig:Optimum r M} depicts the error indicator-based estimates of $r$ and $M$ for separated representation of $OH$ concentration as a function of time. We observe that for most time instances these estimates are $r=1$ and $M=2$, respectively. Notice that these values depend on the number $N$ and particular sets of realizations used to reconstruct the solution. 


\begin{figure}
    \centering
    \begin{tabular}{c}
            \hspace{-0.5cm}    
      \includegraphics[width=3.5in]{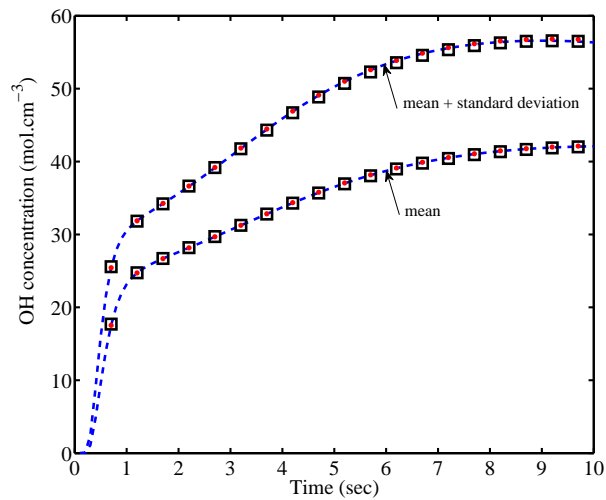}     
          \\
     \end{tabular}
      \caption{ The $OH$ concentration as a function of time. (PC-based reference solution constructed with $N=100,000$ ($---$); PC regression with $N=650$ ($\bullet$); Separated representation with $N=300$ ({\scriptsize $\square$})).}            
 \label{fig: OH profile}       
       \end{figure}

\begin{figure}
    \centering
    \begin{tabular}{c}
            \hspace{-0.5cm}    
      \includegraphics[width=2.9in]{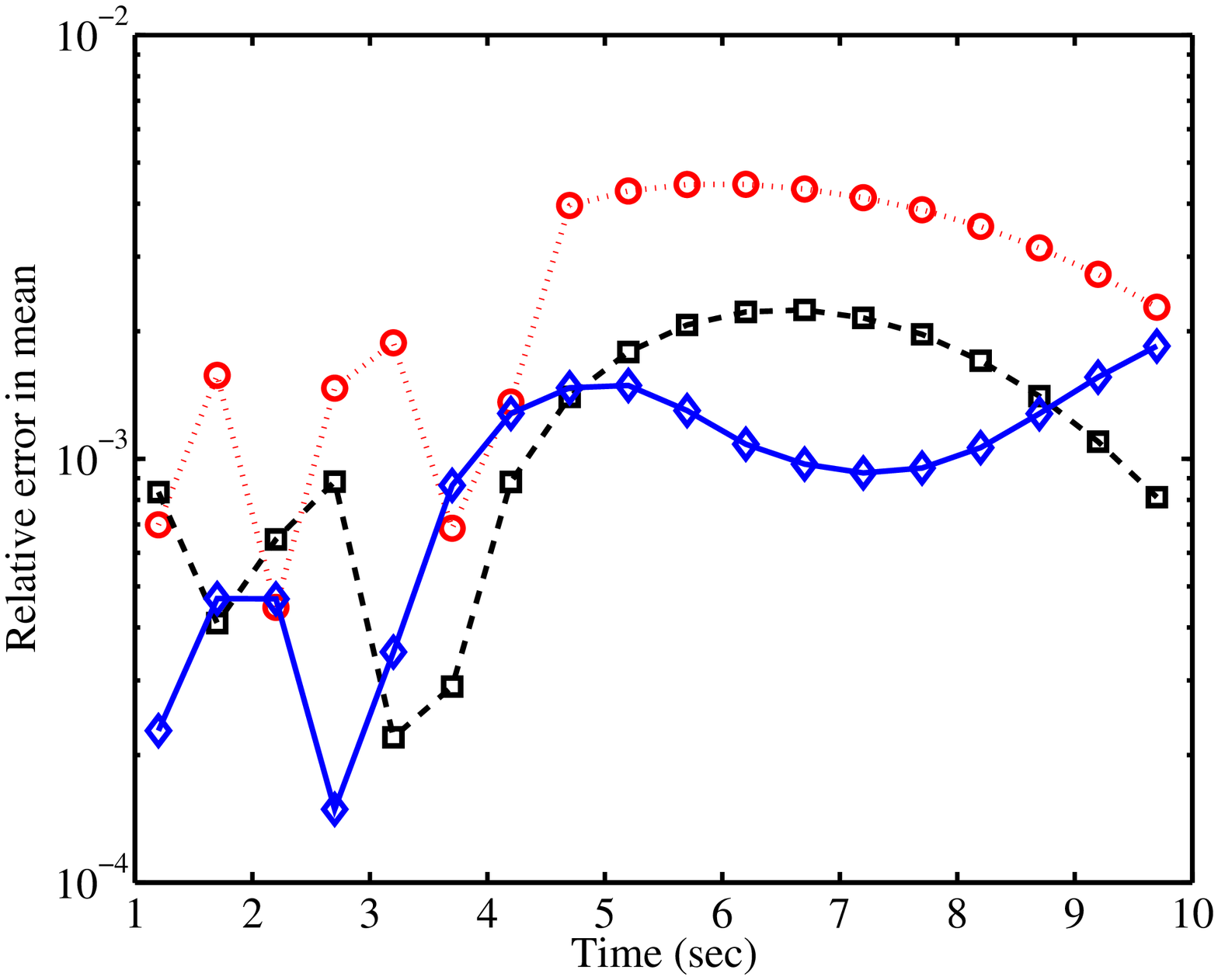}
     
      \hspace{-.5cm}
      \includegraphics[width=2.9in]{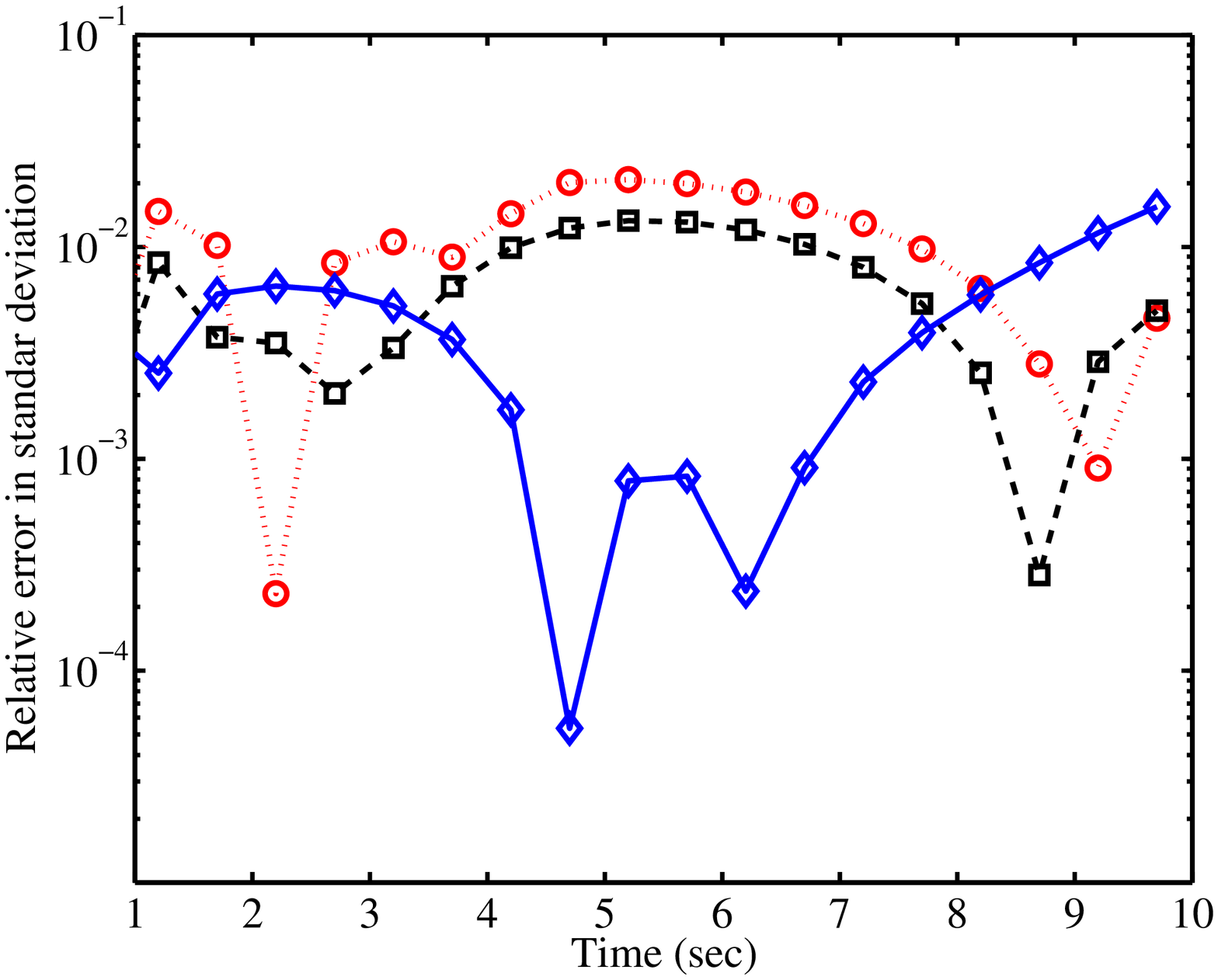}      
      \\
       \put(100,0){\footnotesize(a)}
        \put(300,0){\footnotesize(b)}
          \\
     \end{tabular}
      \caption{Comparison of separated representation and PC regression in estimation of mean and standard deviation of $OH$ concentration. (a) Relative error in mean and (b) Relative error in standard deviation. Separated representation with $N=300$ achieves a similar maximum error (over time $t$) as that of the PC regression. (PC regression with $N=650$ ($\diamond$); Separated representation with $N=300$ ({\scriptsize $\square$});  Separated representation with $N=200$ ($\circ$)).}            
 \label{fig: compare errors for N=200 and N=300 to regression}      
       \end{figure}

We next compare the accuracy of the separated representation and the regression-based PC expansion in estimating the mean and standard deviation of $OH$ concentration, see  Figs. \ref{fig: OH profile} and \ref{fig: compare errors for N=200 and N=300 to regression}. To achieve a stable approximation, the latter approach requires about $N=650$ realizations of $OH$ concentration for a total degree $p=3$ Hermite PC expansion. The accuracy of the $p=2$ PC expansion was found to be significantly below those of the separated representation; therefore, that case was not considered further.  As can be observed from Fig. \ref{fig: compare errors for N=200 and N=300 to regression}, the two approaches achieve similar accuracies, in terms of maximum relative errors over the analysis time interval, while the separated representation requires only about half as many number of realizations of $OH$ concentration.

%


%
\section{Conclusion}
\label{sec:conclusion}

This work introduced a non-intrusive algorithm based on separated representations to approximate model solutions with high-dimensional random inputs. Separated representations may be thought of as generalizations of matrix Singular Value Decomposition (SVD) to higher order tensors and multivariate functions. In particular, a multivariate function is decomposed into a linear sum of unknown functions that are separated with respect to the inputs. In this study, a regression approach was presented to stably compute the separated representation of high-dimensional stochastic functions using their randomly generated realizations. When the number of separated functions in the representation (known as the separation rank) is independent of the number $d$ of random inputs, a successful approximation may be achieved with a number of solution realizations that depends linearly on $d$. The computational complexity of such an approximation grows quadratically in $d$. Therefore, the proposed framework may drastically reduce the issue of curse-of-dimensionality, a bottleneck for uncertainty quantification of systems with high-dimensional random inputs. One of the challenges for regression-based techniques is the issue of over-fitting, which may lead to numerical instabilies and inaccurate approximations. To tackle this issue for our construction, we adopted a Tikhonov regularization approach in which the Tikhonov matrix was designed to penalize large values of the second moment of the approximate solution. The regularization parameter was estimated using the Generalized Cross Validation (GCV) approach. Furthermore, a perturbation-based error indicator was derived to select the parameters of the separated representation to further prevent over-fitting (as well as under-fitting). 
%

The performance of the proposed method was explored through its application to the reconstruction of a manufactured function as well as the numerical solution of two ODE problems with high-dimensional random inputs. The first ODE problem discussed an elliptic differential equation in which the diffusion coefficient was a function of $d=40$ random variables. It was demonstrated, numerically, that the proposed non-intrusive approach outperformed the standard sparse grid stochastic collocation, a widely used technique for approximating high-dimensional functions. The second ODE example focused on quantifying the effect of parametric uncertainties in a hydrogen oxidation problem. The advantage of the proposed approach compared to the regression-based polynomial chaos expansion was observed numerically.

The present construction of separated representation was for scalar-valued stochastic functions. Extensions to vector-valued cases is the subject of an ongoing work. Such an extension is envisioned to exploit the low-rank structure of solution (if exists) in the physical/temporal variables to further reduce the number of required solution realizations.  

\section*{Acknowledgements}
The authors would like to gratefully thank Prof. Gregory Beylkin (CU Boulder) and Prof. Luis Tenorio (Colorado School of Mines) for their valuable suggestions regarding separated representations and regularization of inverse problems.

The work of AD and AV was partially supported by the Department of Energy under Advanced Scientific Computing Research Early Career Research Award DE- SC0006402 and the Predictive Science Academic Alliance Program (PSAAP) at Stanford University. GI gratefully acknowledges financial support from KAUST under award AEA 48803.

\bibliographystyle{plain}
\bibliography{AD_bib_v1}

\end{document}